\documentclass[fleqn,usenatbib]{mnras}

\usepackage{newtxtext,newtxmath}
\usepackage[T1]{fontenc}
\usepackage{ae,aecompl}

\usepackage{graphicx}	% Including figure files
\usepackage{amsmath}	% Advanced maths commands
\usepackage{amssymb}	% Extra maths symbols
\usepackage{tabularx}

\title[Molecular clouds under the FUV feedback]
{Factories of CO-dark gas: molecular clouds with limited star formation efficiencies by FUV feedback}

\newcommand{\sgcl}{\Sigma_{\rm cl}}
\newcommand{\sguni}{{\rm M}_\odot {\rm pc}^{-2}}
\newcommand{\msun}{{\rm M}_\odot}

\author[M. Inoguchi et al.]{Mutsuko Inoguchi$^{1}$\thanks{Contact e-mail: \href{mailto:mutsuko@kusastro.kyoto-u.ac.jp}{mutsuko@kusastro.kyoto-u.ac.jp}},
Takashi Hosokawa$^2$, 
Shin Mineshige$^1$,
and Jeong-Gyu Kim$^3$
\\
$^{1}$Department of Astronomy, Kyoto University, Sakyo-ku, Kyoto 606-8502, Japan\\
$^2$Department of Physics, Kyoto University, Sakyo-ku, Kyoto 606-8502, Japan\\
$^3$Department of Astrophysical Sciences, Princeton University, Princeton, NJ 08544, USA
}

\date{Accepted XXX. Received YYY; in original form ZZZ}

\pubyear{2019}

\begin{document}
\label{firstpage}
\pagerange{\pageref{firstpage}--\pageref{lastpage}}
\maketitle

\begin{abstract}
The star formation in molecular clouds is inefficient. 
The ionizing EUV radiation ($h \nu \geq 13.6$~eV) from young clusters has been considered as a primary feedback effect to limit the star formation efficiency (SFE). We here focus on effects of the stellar FUV radiation (6~eV $\leq h \nu \leq$ 13.6~eV) during the cloud disruption stage. The FUV radiation may further reduce the SFE via photoelectric heating, and it also affects the chemical states of the gas that is not converted to stars ("cloud remnants") via photodissociation of molecules. We have developed a one-dimensional semi-analytic model which follows the evolution of both the thermal and chemical structure of a photodissociation region (PDR) during the dynamical expansion of an HII region. 
We investigate how the FUV feedback limits the SFE, supposing that the star formation is quenched in the PDR where the temperature is above a threshold value (e.g., 100~K).
Our model predicts that the FUV feedback contributes to reduce the SFEs for the massive ($M_{\rm cl} \gtrsim 10^5~{\rm M}_\odot$) clouds with the low surface densities ($\sgcl \lesssim 100~\sguni$). Moreover, we show that a large part of the H$_2$ molecular gas contained in the cloud remnants should be "CO-dark" under the FUV feedback for a wide range of cloud properties. Therefore, the dispersed molecular clouds are potential factories of the CO-dark gas, which returns into the cycle of the interstellar medium. 
\end{abstract}


%%%%%%%%%%%%%%%%%%%%%%%%%%%%%%%%%%%%%%%%%%%%%%%%%%

%%%%%%%%%%%%%%%%% Introduction %%%%%%%%%%%%%%%%%%%

\section{Introduction}
\label{sec:intro}

The evolution of galaxies is closely related to the star formation activities.
In nearby galaxies, the overall star formation rate is quite low; the cold molecular gas is converted to stars in a slow pace. The resulting depletion timescale of the molecular gas is $\sim$ Gyr over the galactic scale \citep[e.g.,][]{Kennicutt12}.  By contrast, the star formation occurs in the much shorter timescale over the small ($\lesssim 100$~pc) cloud scales \citep[e.g.][]{Lee16}. The lifetime of an individual giant molecular cloud (GMC) is estimated to be less than $\sim 10 - 30$~Myr \citep[e.g.,][]{Fukui10}.
A possible explanation for the above observations is that only a small fraction of the gas is used to form stars in each cloud. 
Physical processes responsible for such a low star formation efficiency (SFE) are yet to be fully clarified.  

%-----------------------------------------------------------------------%

A promising process to limit the SFE is the so-called "stellar feedback", i.e., radiative and kinetic energy injection from stars into natal clouds  \citep[e.g.][]{Dale15,Naab17}. The SFEs are lowered if the clouds are promptly destroyed by the feedback before a large part of the gas turns into stars.
Recent studies show that such an evolution is caused indeed by the feedback from high-mass stars in GMCs \citep[e.g.,][]{Kruijssen19}. It is further suggested that the cloud destruction advances over only a few Myrs, which is shorter than the stellar lifetime. Therefore, radiation-driven or wind-driven bubbles expanding around high-mass stars before the first supernova explosion are believed to play an important role in GMC destruction.

%---------------------------------------------------------------------------%

H~II regions created by the stellar ionizing (EUV; $h \nu \geq 13.6$~eV) radiation cause the dynamical bubble expansion in GMCs \citep[e.g.,][]{Yorke86}. 
Since the expansion speed is supersonic with respect to the surrounding cold medium, the H~II bubble expands driving a preceding shock front. The shocked gas is taken into a shell around the H~II region, which continues to expand sweeping up the surrounding medium into the shell. 
While the gas dynamics varies depending on density structure of the clouds \citep[e.g.,][]{Franco90}, theoretical studies have suggested that the resulting EUV feedback operates to limit the SFE \citep[e.g.][]{Whitworth79,Williams97,Matzner02,Kim16}. In recent years, a number of authors have conducted radiation-hydrodynamics numerical simulations that directly follow the EUV feedback in GMCs that are filled with turbulence in reality \citep[e.g.,][]{Mellema06,Dale12,Geen15,Howard16,Gavagnin17,Kim18,Haid19,He19,Gonzalez20}. Overall, these studies have confirmed that the EUV feedback lowers the SFEs, though its impact depends on cloud properties such as the mass and surface density.

%------------------------------------------------------------------------%

%FUV feedback
There are extensive studies regarding the ionizing radiation feedback that have been performed. 
In fact, however, the dissociating radiation (FUV; $6.0~{\rm eV} \leq h\nu \leq 13.6~{\rm eV}$) 
as well as ionizing radiation is emitted from young star clusters.
Many studies have investigated the dynamical effect of the radiation pressure of the FUV radiation. Indeed, some of them shows that the feedback caused by the radiation force contributes to regulating the star formation in GMCs, although the resulting SFE is a bit higher than what observations suggest \citep[e.g.,][]{Thompson2016,Raskutti2016,Raskutti2017,Kim18}.

Whereas the radiation pressure force is one dynamical aspect of the FUV feedback, we in this paper consider the other aspect of the thermal and chemical FUV feedback on GMCs. Hereafter we use the terms of the EUV feedback and FUV feedback to designate the dynamical effects caused by H~II regions and thermal and chemical effects caused by photodissociation region (PDRs), respectively. 
The FUV radiation creates a PDR, where the gas is heated up via photoelectric effect, around an H~II region \citep[e.g.,][]{Hollenbach99}. The local Jeans mass is enhanced by such additional heating, which prevents the gravitational collapse of dense cores. As a result, the FUV radiation may further contribute to the reduction of the SFEs in GMCs \citep[e.g.,][]{RD92,DMF98,Inutsuka+15}. For example, \citet{Forbes16} shows that the photoelectric heating plays the dominant role on determining the star formation rate in dwarf galaxies ($\sim$~kpc-scale) rather than other feedback effects \citep[but see also][]{Hu17}. In the same vein, \citet{Peters17} and \citet{Butler17} have incorporated the FUV feedback in simulations following the star formation in a $\sim$ kpc-scale region of the Galactic disk, concluding that it is necessary to explain the observed depletion timescale of $\sim$ Gyr. On the individual GMC scale ($\lesssim 100$~pc), by contrast, the effects of FUV feedback has not been fully investigated yet.

%-------------------------------------------------------------------%

The low SFE means that most of the GMC gas is returned into the cycle of the insterstellar medium, without being turned into stars. A part of such a "remnant" gas may be recycled for the subsequent GMC formation. 
The stellar FUV radiation also substantially affects the physical and chemical states of the cloud remnants. Since the FUV radiation destroys molecules via photodissociation, it generally creates cold H$_2$ gas associated with little amount of CO molecules \citep[e.g.,][]{vanDishoeck88,Wolfire10}. 
Since such gas is difficult to be observed via CO emission, it is called as "CO-dark" \citep{Dishoeck92}.
Recent observations via $\gamma$-ray \citep[][]{Gre05}, dust continuum \citep[][]{Planck11}, and C$^+$ line emission \citep[][]{Pineda13,Pineda14} suggest the existence of the CO-dark gas, and no less than $\sim$ 30 -- 70~\% of the molecular gas is actually CO dark in our Galaxy. Theoretical studies also support such Galactic-scale observations \citep[e.g.][]{Smith14,Gong18}.
On the cloud scale, the appearance of the CO-dark gas during the formation of molecular clouds has been suggested \citep[e.g.,][]{Clark12}. However, the CO-dark gas phase while the clouds are being dispersed is yet to be further studied \citep[e.g.,][]{Hosokawa07,Gaches2018,Seifried19}. 

%----------------------------------------------------------------------------------%

%our study
As seen above the stellar FUV radiation should cause the additional feedback that affects the SFEs and chemical compositions of the cloud remnants. Whereas fully considering such effects requires expensive numerical simulations of radiation-magneto-hydrodynamics, we here adopt a one-dimensional semi-analytic treatment that guides our understanding. \citet{Kim16} have developed a semi-analytic model for expansion of an H~II region driven by photoionization and radiation pressure. The minimum SFEs limited by the EUV feedback have been evaluated as functions of the cloud masses and surface densities. 
However, they ignore the roles of the FUV feedback. Hence we construct a new model based on \citet{Kim16}, where both the FUV and EUV feedback effects are included. In order to evaluate the FUV feedback, we solve the thermal and chemical structure of PDRs around H~II regions assuming the spherical symmetry. Although simple, this approach allows us to consider impacts of the FUV feedback against a variety of GMCs having different properties. We first investigate how much the FUV feedback contributes to reducing the SFEs. Next we consider the chemical compositions of the GMC remnants under the FUV feedback, showing that they are potential factories of the CO-dark molecular gas.

%-------------------------------------------------------------------------------------%

The rest of the paper is organized as follows. We present our models in Section~\ref{sec:model}, where we outline the overall methodology in Section~\ref{ssec:method} and describe how to couple the dynamics and the thermal and chemical processes operating in the PDR in Sections~\ref{ssec:HIIdyn} and \ref{ssec:strct}. In Section~\ref{sec:result} we show our main results. First we present a representative case of the time evolution of the thermal and chemical structure in the PDR in Section~\ref{ssec:evo}. Then we study the effects of the FUV feedback on limiting the SFE in Section~\ref{ssec:SFE}, and possible chemical compositions of the cloud remnants in Section~\ref{ssec:rem}. We provide the relevant discussion and conclusion in Sections~\ref{sec:dis} and \ref{sec:conc}.

%%%%%%%%%%%%%%%%%%%%%%%%%%%%%%%%%%%%%%%%%%%%%%%%%%

%%%%%%%%%%%%%%%%%%%% Method %%%%%%%%%%%%%%%%%%%%%%

\section{Model}
\label{sec:model}

%%%%%%%%%%%%%%%%%%%%%%%%%%%%%%%%%%%%%%%%%%%%%%%%%%

%%%%%%%%%%%%%%%%%% Methodology %%%%%%%%%%%%%%%%%%%

\subsection{Methodology}
\label{ssec:method}

We first describe our model in this section. We consider spherical and uniform density clouds 
which are characterized by the mass $M_{\rm cl}$ and surface density $\Sigma_{\rm cl}$. 
The cloud radius $R_{\rm cl}$ and hydrogen number density $n_0$ are related to $M_{\rm cl}$ and $\Sigma_{\rm cl}$ as
\begin{equation}
	R_{\rm cl} = \sqrt{M_{\rm cl}/\pi \Sigma_{\rm cl}} 
\label{eq:rcl}
\end{equation}
\begin{align}
	n_0 & = \frac{M_{\rm cl}}{\frac{4}{3} \pi R_{\rm cl}^3 \mu_{\rm H}} \nonumber \\
	& = \frac{3 \pi^{1/2}}{4 \mu_{\rm H}} M_{\rm cl}^{-1/2} \Sigma_{\rm cl}^{3/2} ,
\label{eq:dens}
\end{align}
where $\mu_{\rm H} = 1.4 m_{\rm H}$ is the mean molecular weight per hydrogen nuclei. We consider clouds with $M_{\rm cl} = 10^4,\,10^5,\,10^6$~M$_{\odot}$ below. The surface density is varied so that the resulting number density should fall on a typical range of observed molecular clouds, $30~{\rm cm}^{-3} < n_0 < 10^4~{\rm cm}^{-3}$ \citep[e.g.][]{Tan14}. Table \ref{tab:par} summarizes the ranges of the parameter values we consider. 

\begin{table}
\begin{center}
	\caption{Parameter set}
	\begin{tabular}{llll}\hline \hline
		$M_{\rm cl}$ (M$_{\odot}$) & $\Sigma_{\rm cl}$  (M$_{\odot}/$pc$^2$)  & $R_{\rm cl}$~(pc) & $n_0$~(cm$^{-3}$) \\ \hline
		10$^4$ & 15 -- 700 & 13.2 -- 1.94 & 30 -- 10000\\
		10$^5$ & 31 -- 1506  & 31.9 -- 4.60 & 30 -- 10000\\
		10$^6$ & 67 -- 3246  & 68.7 -- 9.90 &  30 -- 10000\\ \hline	
	\end{tabular}
	\label{tab:par}
\end{center}
\end{table}

%----------------------------------------------------------------------------%

Our aim is to derive minimum SFE required for cloud disruption $\varepsilon_{\rm min}$ as functions of the cloud mass $M_{\rm cl}$ and surface density $\Sigma_{\rm cl}$.% While \cite{Kim16} have considered only the EUV feedback, we here additionally include the FUV feedback to limit the SFE. 
We here focus on the FUV feedback to limit the SFE.
For a given set of $(M_{\rm cl}, \Sigma_{\rm cl})$, we start our calculation by putting a star cluster with the mass of $M_{\ast} = \varepsilon M_{\rm cl}$ at the origin. Here we first take a trial value for the SFE $\varepsilon$. We envision that an H~II region and surrounding photodissociation region (PDR) created by the stellar EUV and FUV radiation expands around the central cluster in the cloud. Following \cite{Kim16}, we calculate the EUV photon number luminosity as
\begin{equation}
S_{\rm EUV} = \Xi_{\rm EUV} M_{\ast}, 
\end{equation}
where the ratio of the stellar mass to the EUV luminosity $\Xi_{\rm EUV}$ is calculated with the SLUG code \citep{K15}. Similarly, we calculate the FUV photon number luminosity
\begin{equation}
S_{\rm FUV} = \Xi_{\rm FUV} M_{\ast}, 
\end{equation}
where we again use the SLUG code to evaluate $\Xi_{\rm FUV}$ (see Appendix \ref{app:ML} for details). We assume that $\Xi$ is time-independent. This is a reasonable approximation, since the dynamical timescale $t_{R_{\rm cl}}$ is shorter than the lifetimes of massive main-sequence stars.
The dynamics of the expanding H~II region and surrounding shell can be described by the analytic formula (see Section \ref{ssec:HIIdyn}). The effects of the FUV radiation on the thermal and chemical structure outside the H~II region are then calculated (Section \ref{ssec:strct}).

%-----------------------------------------------------------------------------%

These calculations are performed using the arbitrary choice of $\varepsilon$, and we determine the minimum SFE by the following iterative procedure. If $\varepsilon$ first assumed is too small, only a small central part of the cloud is affected by the cluster radiation. The further star formation is possible for such a case, meaning that the minimum SFE should be higher. 
We repeat the calculations with increasing $\varepsilon$ incrementally.  
If $\varepsilon$ becomes sufficiently large, the radiative feedback influences the whole natal cloud leaving no room for the further star formation. 
We assume that the minimum SFE $\varepsilon_{\rm min}$ is determined for such a case (section \ref{ssec:cri}). The obtained value of $\varepsilon_{\rm min}$ depends on the feedback effects considered. The FUV feedback potentially reduces the SFE in addition to the EUV feedback because it heats the gas outside the H~II region to hinder the star formation.
The above procedure is basically the same as in \cite{Kim16}, except that we additionally consider the stellar FUV radiation.

%-----------------------------------------------------------------------------%

%%%%%%%%%%%%%%%%%%%%%%%%%%%%%%%%%%%%%%%%%%%%%%%%%%

%%%%%%%%%%%%%%%%%%%% Dynamics %%%%%%%%%%%%%%%%%%%%

\subsection{Dynamics of expanding H II regions}
\label{ssec:HIIdyn}

We here model the dynamical expansion of an H~II region created around the cluster in the natal molecular cloud. In what follows we assume that the photoionized gas has the constant temperature $T_{\rm H\,II} = 10^4$~K for simplicity. The initial size of an H~II region is determined by the so-called Str\"{o}mgren radius 
\begin{equation}
r_{\rm IF,0} = \left( \frac{3 S_{\rm EUV} f_{\rm ion}}{4 \pi n_0^2 (1 - \varepsilon)^2 \alpha_{\rm B}} \right)^{1/3},
	\label{eq:rSt0}
\end{equation}
where $\alpha_{\rm B} = 2.59 \times 10^{-13} (T_{\rm H\,II}/10^4~{\rm K})^{-0.7}$~cm$^3$s$^{-1}$ is 
the case B recombination coefficient \citep[][]{O89}, and $f_{\rm ion} = 0.73$ denotes the fraction of the EUV photons absorbed by the gas \citep[not by the dust, ][]{KM09}.
%%%
We note that $f_{\rm ion}$ varies with the product $S_{\rm EUV} n_{\rm H~II}$ \citep{Draine11}, although the thermal pressure force and the H~II region size only weakly depend on $f_{\rm ion}$ as $r_{\rm IF} \propto f_{\rm ion}^{1/3}$ and $F_{\rm thm} \propto f_{\rm ion}^{1/2}$ \citep{Kim16}.

%-----------------------------------------------------------------------------%
 
Because the internal thermal pressure is much higher than that in the ambient medium, 
the H~II region starts to expand. As considered in \cite{Kim16}, however, the dynamics of the H~II region is generally affected by additional effects such as the radiation pressure exerted on the photoionized gas \citep[e.g.,][]{Draine11} and swept-up shell \citep[e.g.,][]{KM09,Ishiki17}. However, we omit such additional effects for simplicity. Recent theoretical studies show that the radiation pressure effect is particularly important for disrupting GMCs with high surface densities $\sgcl \gtrsim 100~\sguni$ \citep[e.g.,][]{Murray10,Fall10}. We separately examine its effects on our results in Section~\ref{ssec:otherfb}.

%------------------------------------------------------------------------------%

Once the H~II region begins to expand, the ambient gas is swept up to be retained in a shell. The shell is bounded by the ionization front and preceding shock front. The shell mass $M_{\rm sh}$ is estimated as
\begin{equation}
	M_{\rm sh} = \frac{4}{3} \pi r_{\rm IF}^3 \rho_0 (1 - \varepsilon) - M_{\rm H\,II}.
	\label{eq:Msh}
\end{equation}
Here, $r_{\rm IF}$ is the radial position of the ionization front, $M_{\rm H\,II}$ is the mass of ionized gas,
\begin{align}
	M_{\rm H\,II} \approx \frac{4 \pi}{3} r_{\rm IF}^3 \mu_{\rm H} n_{\rm H\,II},
\end{align}
where the
number density of ionized gas $n_{\rm H~II}$ varies with ionization front radius as $n_{\rm H~II} \propto r_{\rm IF}^{-3/2}$. The expansion law, or the time evolution of $r_{\rm IF}$, is derived with the equation of motion of the shell, 
\begin{equation}
	\frac{d}{dt} (M_{\rm sh} v_{\rm sh}) = F_{\rm out} - F_{\rm in},
	\label{eq:shellEOM}
\end{equation}
where $v_{\rm sh} = d r_{\rm sh}/dt$ is the shock velocity, $F_{\rm out}$ and $F_{\rm in}$ represent the forces exerted on the outer and inner surface of the shell. As noted above, we only consider the thermal pressure of the ionized gas as the outward force $F_{\rm out}$,
\begin{equation}
	F_{\rm thm} = 4 \pi r_{\rm IF}^2 \cdot 2 n_{\rm H\,II} k_{\rm B} T_{\rm H\,II},
\end{equation}
which scales as $F_{\rm thm} \propto n_{\rm H~II} r_{\rm IF}^2 \propto r_{\rm IF}^{1/2}$.  We ignore $F_{\rm in}$ for simplicity. Equation (\ref{eq:shellEOM}) is solved analytically, and we obtain
\begin{equation}
	r_{\rm IF} (t) = r_{\rm IF,0} \left( 1 + \frac{7}{4} \sqrt{\frac{4}{3}} \frac{c_{\rm s} t}{r_{\rm IF,0}} \right)^{4/7},
	\label{eq:rsh}
\end{equation}
where $c_{\rm s} = \sqrt{ 2 k_{\rm B} T_{\rm H\,II}/\mu_{\rm H}}$ is the sound speed in H II region \citep[][]{HI06}. Equation (\ref{eq:rsh}) differs from the well-known expansion law given by \citet{Spitzer78} by the factor of $\sqrt{4/3}$, but it actually provides the better approximation as proven by radiation-hydrodynamics numerical simulations \citep[e.g.,][]{Bisbas15,Kim2017,Williams18}. Note that equation (\ref{eq:rsh}) is basically the same as that given by \cite{Kim16} but we only consider the thermal pressure of the photoionized gas. 
\cite{Haworth2015} performed RHD simulations of expanding H~II region by taking into account of microphysics such as detailed thermal processes and chemistry. They showed that the expansion is slightly delayed by the order of 10~\% at most. It is reasonable to use equation (\ref{eq:rsh}) in our calculation.

%%%%%%%%%%%%%%%%%%%%%%%%%%%%%%%%%%%%%%%%%%%%%%%%%%

%%%%%%%%%%%%%%%%%%% Structure %%%%%%%%%%%%%%%%%%%%

\subsection{Thermal and chemical structure of photodissociation regions}
\label{ssec:strct}

For every snapshot of an expanding H~II region within the cloud, we calculate the thermal and chemical structure in the surrounding photodissociation region (PDR). Below we consider the following seven chemical species of e$^-$, H$^+$, H$^0$, H$_2$, C$^+$, O$^0$ and CO. We assume the total abundance of C and O atoms as $x_{\rm C} = 3.0 \times 10^{-4}$ and $x_{\rm O} = 4.6 \times 10^{-4}$ \citep[][]{W95}, where $x$ denotes the number fraction relative to the hydrogen nuclei.

%%%%%%%%%%%%%%%%%%%%%%%%%%%%%%%%%%%%%%%%%%%%%%%%%%

%%%%%%%%%%%%%%%%%%%% One-zone %%%%%%%%%%%%%%%%%%%%

\subsubsection{One-zone thermal and chemical equilibrium model}
\label{sssec:one-zone}

\newcolumntype{C}{>{\centering\arraybackslash}X}
\newcolumntype{R}{>{\raggedright\arraybackslash}X}

\begin{table*}
	\begin{center}
	\caption{The thermal and chemical processes included in our model}
	\begin{tabularx}{1.5\columnwidth}{p{32mm}p{70mm}C} \hline \hline
	& Processes &  Reference \\ \hline
	Heating $\Gamma (n, T, x^{\ast})$ & photoelectric heating & 1 \\
	& ionization by soft X-ray & 2 \\
	& H$_2$ photodissociation & 3 \\
	& H$_2$ formation & 3 \\
	Cooling $\Lambda (n, T, x^{\ast})$ & fine structure line emission  & \\
	& \hspace{1mm} [C II] 158~$\mu$m & 3 \\ 
	& \hspace{1mm} [O I] 63~$\mu$m, 44.2~$\mu$m, 145.6~$\mu$m & 3\\
	& Ly $\alpha$ line emission  & 4\\
	& CO rotational line emission & 5\\
	& collision with dust grains & 6\\ \hline
	$R^{\rm H^+}_{\rm form} (n, T, x^{\ast})$ & ionization by soft X-ray & 2\\
	$R^{\rm H^+}_{\rm rec} (n, T, x^{\ast})$ & case B recombination & 7 \\ \hline
	$R^{\rm H_2}_{\rm form} (n, T, x^{\ast})$ & dust catalysis & 8\\
	& associative detachment  & 3\\
	$R^{\rm H_2}_{\rm dis} (n, T, x^{\ast})$ & photodissociation & 8,9\\
	& dust collision & 8\\ \hline
	$R^{\rm CO}_{\rm form} (n, T, x^{\ast})$ & CO formation & 10, 11\\
	$R^{\rm CO}_{\rm dis} (n, T, x^{\ast})$ & photodissociation & 10, 11\\ \hline
	\multicolumn{3}{p{1.45\columnwidth}}{References:
	(1) \citet{BT94}; 
	(2) \citet{W95}; 
	(3) \citet{HM79};
	(4) \citet{Spitzer78}; 
	(5) \citet{Mckee82}; 
	(6) \citet{HM89}; 
	(7) \citet{O89}; 
	(8) \citet{TH85}; 
	(9) \citet{Draine96};
	(10) \citet{L76}; 
	(11) \citet{NL97}
	} \\
	\end{tabularx}
	
	\label{tab:physpro}
\end{center}
\end{table*}

We make use of the one-zone modeling of the thermal and chemical equilibrium state of the interstellar medium \citep[e.g.,][]{W95,KI00}. Consider the gas with a given density $n$ exposed by a FUV radiation field with $G_0$. We determine the unknown variables, the gas temperature $T$ and chemical number fractions $x_{\rm H^+}, x_{\rm H_2}, x_{\rm CO}$, by solving the following equations 
\begin{align}
	&\frac{d e}{dt} = \Gamma (n, T, x^{\ast}) - \Lambda (n, T, x^{\ast})	\label{eq:thm} \\
	&\frac{dx_{\rm H^+}}{dt} = R_{\rm H^+}^{\rm form} (n, T, x^{\ast}) - R_{\rm H^+}^{\rm rec} (n, T, x^{\ast})\\	
%	\label{eq:H+}
	&\frac{dx_{\rm H_2}}{dt} = R_{\rm H_2}^{\rm form} (n, T, x^{\ast}) - R_{\rm H_2}^{\rm dis} (n, T, x^{\ast}) \label{eq:H2} \\
	&\frac{dx_{\rm CO}}{dt} = R_{\rm CO}^{\rm form} (n, T, x^{\ast}) - R_{\rm CO}^{\rm dis} (n, T, x^{\ast})
	\label{eq:CO}
\end{align}
where $e$ is internal energy of the gas,   $\Gamma$ and $\Lambda$ are the heating and cooling rates, and $x^{\ast}$ represents $(x_{\rm H^+}, x_{\rm H_2}, x_{\rm CO})$.
In the present study, we only consider C$^+$ and CO as carbon compounds and thus set $x_{\rm C^+} = x_{\rm C} - x_{\rm CO}$. 

%------------------------------------------------------------------------------%

A full list of thermal and chemical processes associated with the terms on the R.H.S of equations (\ref{eq:thm}) - (\ref{eq:CO}) is presented in Table \ref{tab:physpro}. We here only briefly describe some of them. Those readers who are interested in more details may refere to the references therein.  
As the heating processes, we incorporate the photoelectric emission from grains and H$_2$ dissociation by the FUV radiation, ionization by the background soft X-ray radiation, and H$_2$ formation releasing the binding energy. The radiative cooling is primarily caused via the line emission of [C II], [O I], Ly-$\alpha$, and CO. We assume the optically-thin limit for these line emission. It is equivalent to ignoring the trapping effect, for which possible effects on our conclusions are discussed in \textcolor{blue}{Section~\ref{ssec:ignored}}. To avoid overcooling, we set the minimum gas temperature to be 8~K. Regarding the formation of CO molecules, we adopt the simple method given by \citet{NL97}, where CO molecules are approximately formed from C$^+$ ions and O atoms. 
\cite{Gong18} pointed out that the \cite{NL99} chemical network significantly underestimates CO abundance for $n \lesssim 500$~cm$^{-3}$ and $A_{\rm V}$ < 5. However, we use the chemical network by Nelson \& Langer in the present study, since we focus on the CO abundance at dense shell where $n > 10^4$~cm$^{-3}$. We also assume the constant dust temperature $T_{\rm d} = 8$~K for the all cases considered. The dust temperature is used to estimate the reformation rate of H$_2$ molecules and the thermal gas-dust coupling rate via collisions. We also investigate the effects of varying $T_{\rm d}$ in our calculations in \textcolor{blue}{Section~\ref{ssec:ignored}}.

%-------------------------------------------------------------------------------%

%%%%%%%%%%%%%%%%%%%%%%%%%%%%%%%%%%%%%%%%%%%%%%%%%%

%%%%%%%%%%%%%%%% Time-evolution %%%%%%%%%%%%%%%%%%

\subsubsection{Time-evolution of multi-zone structure}

%-----------------------------------------------------------------------------%

We calculate the spatial variation of the thermal and chemical state in the PDR around an H~II region by repeating the one-zone calculations as follows. At a given time $t = t_j$, the radius and mass of the shell, $M_{\rm sh} (t_j)$ and $r_{\rm sh} (t_j)$, are described by equations (\ref{eq:Msh}) and (\ref{eq:rsh}). 
By setting radial grids, we discretize the outer PDR including the shell into cells with the column density $\Delta N_{\rm H} \sim 10^{19}~{\rm cm}^{-2}$ per each
which corresponds to $A_V = 5.0 \times 10^{-3}$ with the conversion law of $A_V = 5.0 \times 10^{-22} N_{\rm H}$.
The number of the grids is typically $\sim 1000$.
The distance from the ionization front to the $i$-th grid $r_i$ is 
\begin{equation}
	r_i = r_{\rm IF}(t_j) + \sum_{k = 0}^i \Delta N_{\rm H}/n_k,
\end{equation}
which corresponds to the dust optical depth in the outward direction
\begin{equation}
	\tau_{{\rm in},\,i} = \sigma_d \sum_{k = 0}^i \Delta N_{\rm H},
\end{equation}
and the dust optical depth from the edge of the cloud $\tau_{{\rm out},\,i}$
\begin{equation}
	\tau_{{\rm out},\,i} = \sigma_d \sum_{k = i}^{N} \Delta N_{\rm H},
\end{equation}
The normalized FUV flux at $r= r_i$ is written as
\begin{equation}
	G_i = \frac{1}{F_{\rm H}} \frac{S_{\rm FUV}}{4 \pi r_i^2} \exp (-\tau_{\rm in},\,i) + G_{\rm bg}  \exp(-\tau_{{\rm out},\,i}),
	\label{eq:Gi}
\end{equation}
where $N$ is the total number of the grids, 
 $\sigma_{\rm d} = 10^{-21}$~cm$^2 {\rm H}^{-1}$ is the absorption cross section by dust grains per hydrogen nucleus,  and $F_{\rm H} = 1.21 \times 10^7~{\rm cm}^{-2} {\rm s}^{-1}$ is the normalization factor which represents the background field near the Solar system \citep[the so-called Habing unit, i.e., ][]{Habing68,Draine96} . The last term of the unity in equation (\ref{eq:Gi}) represents this background exactly. The mass summation over the cells located at $r \leq r_i$ is
\begin{equation}
	M_i = \sum_{k = 0}^i 4 \pi r_k^2 \mu_{\rm H} \Delta N_{\rm H}.
\end{equation}
By comparing $M_i$ to the total shell mass $M_{\rm sh}$, we judge whether the $i$-th cell is still within the shell or not. As far as $M_i < M_{\rm sh}$, the cell is regarded as a part of the shell. We determine the thermal and chemical states of such cells in an iterative manner as follows. We assume that the gas pressure within the shell is equal to that of the H~II region, $P_{\rm th} = 2n_{\rm HII} k_{\rm B}T_{\rm HII}$. So we initially provide the pressure instead of the density in a one-zone calculation, unlike in Section~\ref{sssec:one-zone}.
With the given pressure $P_{\rm th}$ and FUV field $G_i$, we calculate the unknown variable $(T_i, x^{\ast}_i)$ by solving equations (\ref{eq:thm})-(\ref{eq:CO}) so that the resulting pressure $P_{\rm sh} = n_i(1+x_{e^-} - x_{{\rm H}_2}/2) k_{\rm B} T_i$ matches $P_{\rm th}$. By doing that, we also determine the number density $n_i$ as well as $(T_i, x^{\ast}_i)$. Once $(n_i, T_i, x^{\ast}_i)$ are fixed, we then move on to the next $(i+1)$-th cell and repeat the same procedures. If $M_i$ exceeds $M_{\rm sh}$, the following cells are considered to be outside of the shell as the un-shocked ambient gas. We take exactly the same method as in Section \ref{sssec:one-zone} for such cells; we calculate $(T_i, x^{\ast}_i)$ for the given number density $n_0 (1-\varepsilon)$ and FUV field $G_i$.
We continue the calculations until reaching the cloud edge, i.e., for $M_i < M_{\rm gas} = M_{\rm cl} (1 - \varepsilon) - M_{\rm H\,II}$.

%%%%%%%%%%%%%%%%%%%%%%%%%%%%%%%%%%%%%%%%%%%%%%%%%%
%%%%%%%%%%%%%%%%%%% Criteria %%%%%%%%%%%%%%%%%%%%%

\subsection{Cloud disruption criteria}
\label{ssec:cri}

To determine the minimum SFE of the cloud, we need some criteria of the cloud disruption as in \cite{Kim16}. We investigate the effects of the FUV feedback on top of the EUV feedback previously studied. So we first use the exactly the same criterion as in \cite{Kim16}:
\\

{\bf Criterion 1 (EUV feedback):} An H~II region and shell are assumed to expand as far as the shell velocity $v_{\rm sh}$ is larger than the critical velocity $v_{\rm bind} = \sqrt{GM_{\rm cl}(1 + \varepsilon)/R_{\rm cl}}$,
\begin{equation}
 v_{\rm bind} \simeq 5~{\rm km/s} \left( \frac{M_{\rm cl}}{10^5~{\rm M}_\odot} \right)^{1/4}
\left( \frac{\sgcl}{10^2~\sguni} \right)^{1/4} (1 + \varepsilon )^{1/2} .
\end{equation}
If the trial value of $\varepsilon$ is too small, the expansion stalls well before the shell reaches the cloud edge. We iteratively increase $\varepsilon$ until $v_{\rm sh} = v_{\rm bind}$ is satisfied at the cloud edge, i.e., $r = R_{\rm cl}$. This gives the minimum SFE.
\\

Note that the above is not the only criterion investigated in \cite{Kim16}. They have also adopted other criteria, showing that the obtained minimum SFE does not largely change. Since our aim is to study the effects of the FUV radiation, we only focus on one representative case. 
\\

{\bf Criterion 2 (FUV feedback):} 
We assume that the star formation is suppressed in a warm PDR, where the gas temperature is above the threshold value $100$~K. Technically, if the trial value of $\varepsilon$ is too small, the temperature outside of the shell is at least  partly lower than 100~K. We iteratively increase $\varepsilon$ until the gas is heated above 100~K everywhere outside the shell at a certain epoch. This gives the minimum SFE.

%------------------------------------------------------------------------------------------%

Although the temperature is raised to $\sim 100-1000$~K in the PDR, the corresponding sound speed is much smaller than that of the photoionized gas. Therefore, as often presumed, the resulting FUV feedback should be weaker than the EUV feedback. The FUV effects would not operate to disrupt the entire structure of the molecular clouds. We rather suppose that the star formation in the PDR is locally hindered with the lack of the cold ($\sim 10$~K) materials. Since the exact strength of the FUV feedback is uncertain, we also consider Criterion 1 for limiting the SFEs. We only estimate effects of the FUV feedback on the chemical compositions of cloud remnants for such cases.

%%%%%%%%%%%%%%%%%%%%%%%%%%%%%%%%%%%%%%%%%%%%%%%%%%

%%%%%%%%%%%%%%%%%%%% Results %%%%%%%%%%%%%%%%%%%%%%

\section{Results}
\label{sec:result}

%%%%%%%%%%%%%%%%%%%%%%%%%%%%%%%%%%%%%%%%%%%%%%%%%%

%%%%%%%%%%%%%%%% Time-evolution %%%%%%%%%%%%%%%%%%

\subsection{Time evolution of thermal and chemical structure}
\label{ssec:evo}

\begin{figure}
	\begin{center}
		\includegraphics{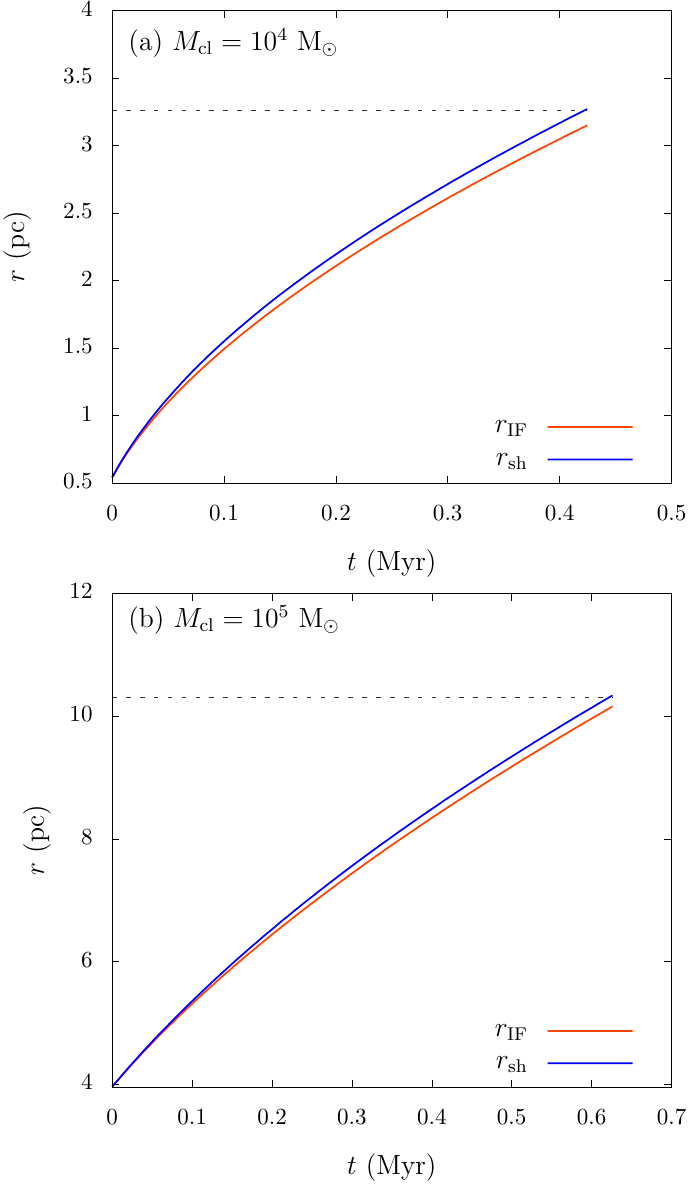}
	\end{center}
	\caption{
The positions of the ionization front $r_{\rm IF}$ (red solid line) and the shock front $r_{\rm sh}$ (blue solid line) as functions of time for the cases with (a) $M_{\rm cl} = 10^4$~M$_{\odot}$ ($\Sigma_{\rm cl} = 300$~M$_{\odot}$pc$^{-2}$)
	and with (b) $M_{\rm cl} = 10^5$~M$_{\odot}$ ($\Sigma_{\rm cl} = 300$~M$_{\odot}$pc$^{-2}$) in the upper and lower panels, respectively.
	The black dashed line in each panel indicates the position of the outer edge of the cloud; (a) $R_{\rm cl} = 3.25$~pc and (b) $R_{\rm cl} = 10.3$~pc.
	}
	\label{fig:t-R_M4}
\end{figure}

\begin{figure*}
	\begin{center}
		\includegraphics[width = 18cm]{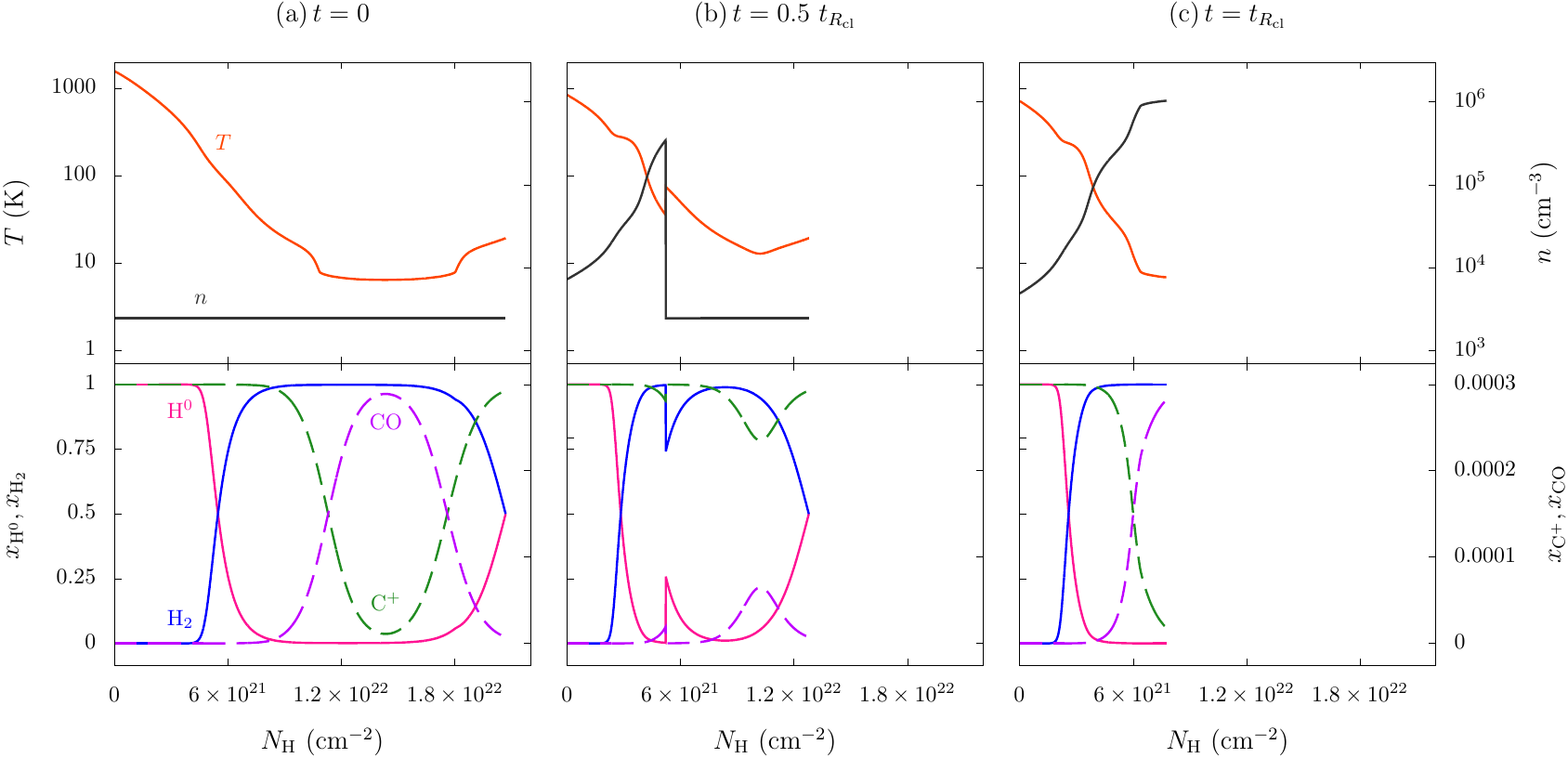}
	\end{center}
	\caption{
	Time evolution of the thermal and chemical structure in the photodissociation region around an H~II region. The cloud mass and surface density are $M_{\rm cl} = 10^4$~M$_{\odot}$ and $\Sigma_{\rm cl} = 300$~M$_{\odot}$pc$^{-2}$ for this case. The panels (a), (b), and (c) show the snapshots at the different epochs of (a) $t = 0$, (b) $t = 0.5 \, t_{R_{\rm cl}}$ and (c) $t = t_{R_{\rm cl}}$, where $t_{R_{\rm cl}}$ is the time when the shell reaches the cloud edge. The horizontal axis denotes the column density of hydrogen nuclei measured from the ionization front; 
	that is, $N_{\rm H} = 0$ corresponds to $r_{\rm IF}$ and the maximum value of $N_{\rm H}$ corresponds to $R_{\rm cl}$.
	{\it Top:} Plotted are the gas temperature (red line) and density (gray line), for which the scaling is presented with the left- and right-hand axis. {\it Bottom:} Plotted are the fractional abundances of H~I (red solid line), H$_2$ (blue solid line), C~II (green dashed line), and CO (purple dashed line). The left-hand (right-hand) axis is used for scaling of H~I and H$_2$ (C~II and CO) abundances.
	}
	\label{fig:NH-TX}
\end{figure*}

\begin{figure*}
	\begin{center}
		\includegraphics[width = 18cm]{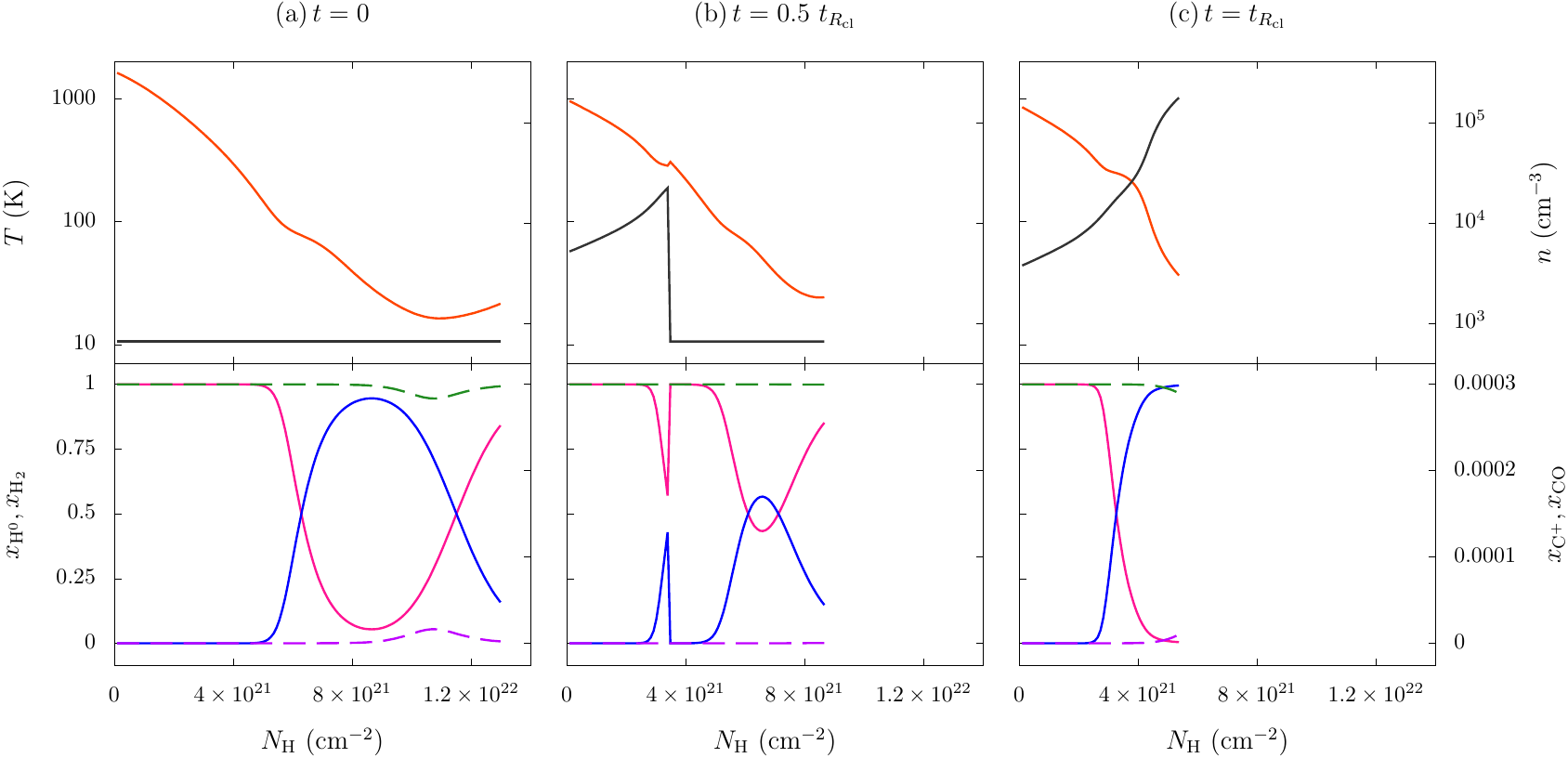}
	\end{center}
	\caption{
	Same as Fig. \ref{fig:NH-TX}, except for the higher cloud mass of $M_{\rm cl} = 10^5$~M$_{\odot}$ and surface density $\sgcl = 300~\sguni$.
	}
	\label{fig:NH-TX_M5}
\end{figure*}

First we present typical evolution of the thermal and chemical structure in the PDR around an H~II region. Here we spotlight one particular case with the molecular cloud mass $M_{\rm cl} = 10^4$~M$_{\odot}$ and surface density $\Sigma_{\rm cl} = 300$~M$_{\odot}$pc$^{-2}$.
We follow the evolution with a star cluster with $M_* = \varepsilon_{\rm min} M_{\rm cl} \simeq 1220~$M$_\odot$ formed at the cloud center\footnote{ For this representive case, we find that the minimum SFE $\varepsilon \simeq 0.12$ is insensitive to the choice of cloud disruption criteria (see also Section~\ref{ssec:SFE}).}. The corresponding stellar EUV and FUV photon number luminosities are $S_{\rm EUV} \simeq 4.2 \times 10^{49}~{\rm sec}^{-1}$ and $S_{\rm FUV} \simeq 8.0 \times 10^{49}~{\rm sec}^{-1}$ respectively.

Fig. \ref{fig:t-R_M4} shows the position of ionization front $r_{\rm IF}$ (equation \ref{eq:rsh}) and shell $r_{\rm sh} = r_{\rm IF} + \sum \Delta N_{\rm H}/n_i$ as a function of time. 
Fig. \ref{fig:NH-TX} shows the time evolution of the one-dimensional thermal and chemical structure 
at (a) $t = 0$, (b) $t = 0.5 \, t_{R_{\rm cl}}$, and (c) $t = t_{R_{\rm cl}}$, 
where $t_{R_{\rm cl}}$ is the time when the shell reaches the cloud edge. 
Note that the total column density deceases with time in this figure.
This is explained by the difference of the geometry: 
the initial and final column density $N_0 = n_0 R_{\rm cl}$ and  
 $N_{\rm H}^{\rm shell}$ are related as 
\begin{align}
	M_{\rm gas} = \frac{4}{3} \pi R_{\rm cl}^2 N_0 \mu_{\rm H} \sim 4 \pi R_{\rm cl}^2 N_{\rm H}^{\rm shell} \mu_{\rm H},
\end{align}
where we approximate $r_{\rm IF} (t_{R_{\rm cl}})$ as $R_{\rm cl}$.
Then we find $N_{\rm H}^{\rm shell} \sim N_0/3$.

%---------------------------------------------------------------------------------%

Fig. \ref{fig:NH-TX}~(a) presents the snapshot at $t=0$, when the initial Str\"{o}mgren sphere is created. Since at this epoch the shell has not appeared yet, the density is constant everywhere. The temperature rapidly grows toward the central cluster because of the efficient photoelectric heating by the strong stellar FUV radiation. 
In the outer part with $1.2 \times 10^{22}$~cm$^{-2} \lesssim N_{\rm H} \lesssim 1.8 \times 10^{22}$~cm$^{-2}$, however, the temperature profile is flat since we set the minimum gas temperature at 8~K (see Section \ref{sssec:one-zone}).
%In the outer part with $N_{\rm H} \gtrsim 10^{22}$~cm$^{-2}$, however, the temperature is almost constant because the FUV flux from the cluster becomes negligible compared to the background field (equation \ref{eq:Gi}).
In the lower panel, we see that the hydrogen molecules are dissociated by the cluster FUV radiation for $N_{\rm H} \lesssim 4.0 \times 10^{21}$~cm$^{-2}$.

%--------------------------------------------------------------------------------%

Fig.~\ref{fig:NH-TX}~(b) shows that the swept-up shell has emerged by the epoch of $t = 0.5 \, t_{R_{\rm cl}}$ and $r_{\rm IF} = 2.2$~pc. The discontinuity of physical quantities at $N_{\rm H} \simeq 5.4 \times 10^{21}$~cm$^{-2}$, which corresponds to the preceding shock front, or the shell outer edge represented by $r_{\rm sh}$. 
Within the shell, the temperature decreases outward as the FUV flux drops owing to the dust attenuation. The density inversely increases, because the thermal pressure is assumed to be fixed at the value of the H~II region. The hydrogen dissociation front is shifted to the lower column density at $N_{\rm H} \simeq 2.5 \times 10^{21}$~cm$^{-2}$ than in panel (a) because of the efficient self-shielding of H$_2$ molecules within the dense shell. By contrast, there is only little amount of CO molecules within the shell. The temperature just outside the shell is slightly higher than that inside the shell
because the [C~II] line emission, which is the dominant coolant of the cloud, is less efficient with the lower density (see also Section \ref{sssec:why}). 
Since the density differs by approximately 2 orders of magnitude across the shock front, the cooling efficiency also differs.

%---------------------------------------------------------------------------------%

Fig.~\ref{fig:NH-TX}~(c) shows the final snapshot for the current case, when all of the cloud materials are swept into the shell. Unlike the previous snapshot, the CO dissociation front is taken into the shell at $N_{\rm H} \simeq 6.0 \times 10^{21}$~cm$^{-2}$ because the shell column density has become so large that CO molecules are protected against the cluster FUV radiation with the dust attenuation. As shown below, this is the final snapshot when the minimum SFE is determined, and the swept-up gas on the shell is, so to speak, the remnant of the molecular cloud. It is evident that the chemical composition of such a cloud remnant is not homogeneous. There are some amount of H$_2$ molecules, but only a small part of those is associated with CO molecules. We return to this point later in Section~\ref{ssec:rem}.

%%--------------------------------------------------------------------%%

Next we show the case where CO molecules are almost completely destroyed by FUV radiation. Fig.~\ref{fig:NH-TX_M5} represents the case with $M_{\rm cl} = 10^5$~M$_{\odot}$ and $\sgcl = 300~\sguni$. The central cluster mass is $2.6 \times 10^4$~M$_{\odot}$ and corresponding stellar EUV and FUV photon number luminosity is $S_{\rm EUV} \simeq 1.2 \times 10^{51}$~s$^{-1}$ and $S_{\rm FUV} \simeq 2.5 \times 10^{51}$~s$^{-1}$, respectively. The clear difference from the case with $M_{\rm cl} = 10^4$~M$_{\odot}$ is that CO molecules do not survive throughout the time evolution. This behavior is mainly explained by the difference of $G_0$ (see Section \ref{ssec:rem} for detailed discussion).

%%%%%%%%%%%%%%%%%%%%%%%%%%%%%%%%%%%%%%%%%%%%%%%%%%

%%%%%%%%%%%%%%%%%% Minimum SFE %%%%%%%%%%%%%%%%%%%

\subsection{Star formation efficiency of molecular clouds}
\label{ssec:SFE}
\subsubsection{Limiting star formation efficiency by FUV radiation}
\label{sssec:SFE}

\begin{figure*}
	\begin{center}
		\includegraphics[width = 18cm]{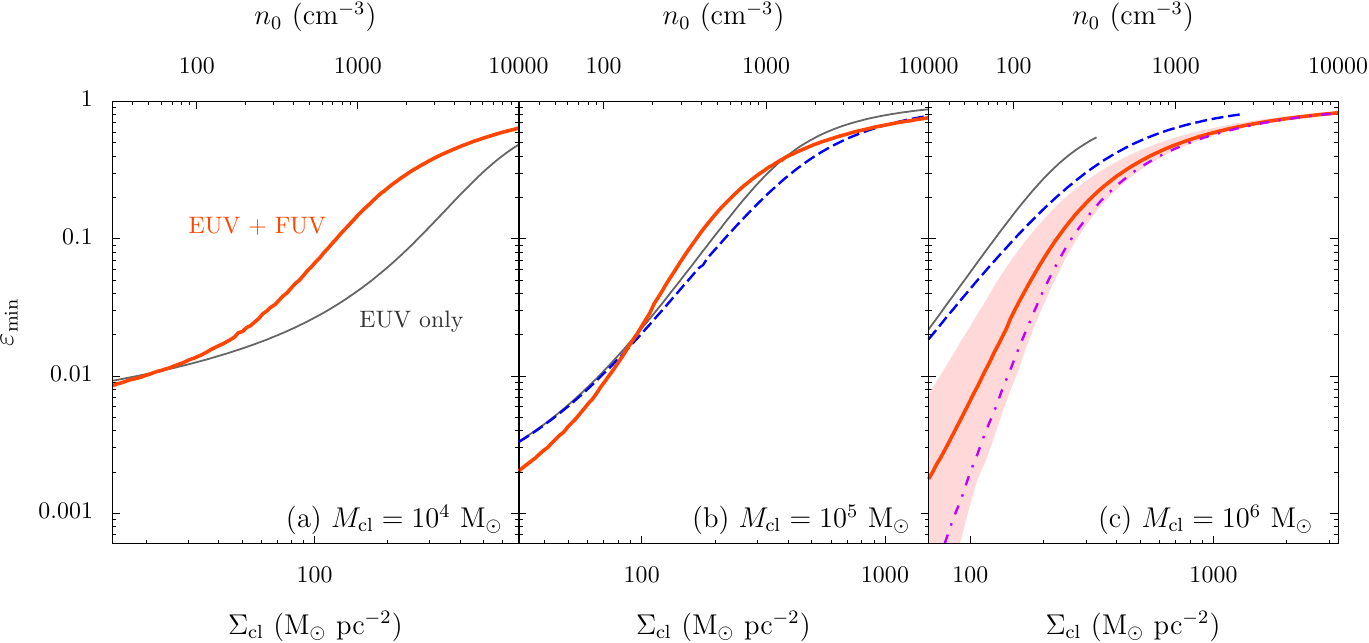}
	\end{center}
	\caption{
	The minimum star formation efficiency (SFE) $\varepsilon_{\rm min}$ calculated as functions of the cloud surface density $\Sigma_{\rm cl}$. 
	The different cloud masses of $M_{\rm cl} = 10^4$~M$_{\odot}$, $M_{\rm cl} = 10^5$~M$_{\odot}$ and $M_{\rm cl} = 10^6$~M$_{\odot}$ are assumed for panels (a), (b) and (c). 
	In each panel, the thick red line represents the case where the SFE is limited by both the EUV and FUV feedback (Criterion 2). The black line represents the reference case only with the EUV feedback (Criterion 1), as considered in \citet{Kim16}. 
	The blue dashed lines in panel (b) and (c) represents the cases where effects of the radiation pressure is included for the dynamics of the H~II region expansion.
	In panel (c), the red shaded zone represents the range where the threshold temperature is varied between 50~K and 300~K in Criterion 2, and the purple dot-dashed line represents the case with lower C and O abundances, $x_{\rm C} = 1.4 \times 10^{-4}$ \citep{Cardelli96} and $x_{\rm O} = 2.8 \times 10^{-4}$ \citep{Cartledge04}.
	Note that each panel shows a different range of $\Sigma_{\rm cl}$.
	}
	\label{fig:SFE}
\end{figure*}

In this section we investigate the SFE of the molecular clouds set by the EUV and FUV feedback effects. Consider an expanding H~II region and surrounding PDR around a newly-born cluster in a given molecular cloud. If the cluster is not sufficiently massive (or luminous), only a small part of the cloud near the cluster would be affected by the feedback; further star formation would occur in the remnant part until enough stars have formed to halt further star formation and destroy the whole cloud. Hence there should be the minimum value of the SFE $\varepsilon_{\rm min}$ above which the cloud is destroyed by radiative feedback. We calculate $\varepsilon_{\rm min}$ as functions of $M_{\rm cl}$ and $\Sigma_{\rm cl}$ in the iterative manner as outlined in Section~\ref{ssec:method}.

%-----------------------------------------------------------------------------------------%

Each panel in Fig.~\ref{fig:SFE} shows the minimum SFE obtained as a function of the cloud surface density $\Sigma_{\rm cl}$ for the same mass $M_{\rm cl}$. The cloud masses of $M_{\rm cl}$ = 10$^4$, 10$^5$ and 10$^6$~M$_{\odot}$ are assumed for panels (a), (b) and (c), respectively. 

The gray line in each panel represents the case where only the EUV feedback is considered (Criterion 1, $\varepsilon_{\rm min,1}$).
We see that $\varepsilon_{\rm min,1}$ is an increasing function of $\Sigma_{\rm cl}$, as shown in \cite{Kim16}. Such a behavior is well understood by considering the $\Sigma_{\rm cl}$-dependences of the cloud radius $R_{\rm cl}$ and initial Str\"{o}mgren radius $r_{\rm St,\,0}$: $R_{\rm cl} \propto \Sigma_{\rm cl}^{-1/2}$ and $r_{\rm St,\,0} \propto \Sigma_{\rm cl}^{-1}$ for a given $M_{\rm cl}$ and $S_{\rm EUV}$.
It means that, with increasing $\Sigma_{\rm cl}$, the typical size of the H~II region $r_{\rm St,\,0}$ relative to the cloud size $R_{\rm cl}$ decreases. The more massive or luminous cluster is necessary for the H~II region to cover the whole cloud for such a case. Thus the resulting $\varepsilon_{\rm min}$ is higher for higher surface density. \citet{Kim16} provide the analytic formula describing this dependence as 
\begin{equation}
	\frac{\varepsilon_{\rm min}}{(1 - \varepsilon_{\rm min}^2)^2} = \left( \frac{\pi^{5/4} G}{\eta_{\rm th} \mathcal{T}} \right)^2 M_{\rm cl}^{1/2} \Sigma_{\rm cl}^{5/2},
	\label{eq:SFE_kim}
\end{equation}
where $\eta_{\rm th} = 9/4$ and 
$\mathcal{T} = 8 \pi k_{\rm B} T_{\rm H\,II} [3 f_{\rm ion} \Xi_{\rm EUV}/4 \pi \alpha_{\rm B}]^{1/2}$.
Note that the gray line in each panel representing the EUV feedback is not identical because of the dependence of $\varepsilon_{\rm min} \propto M_{\rm cl}^{1/2}$ in equation (\ref{eq:SFE_kim}). 

In our model, the gas density is proportional to $(1 - \varepsilon)$ and the photon number flux $S_{\rm EUV}$ is proportional to $\varepsilon$, so that the size of the initial H~II region becomes increasingly larger for higher $\varepsilon$. Thus there is a critical $\varepsilon$ over which $r_{\rm IF,0} \geq R_{\rm cl}$. This occurs when the  cloud surface density and mass are both large \citep[see also][]{Kim16}.
This explains why the gray solid line stops in the middle of the diagram in panel (c).

%-----------------------------------------------------------------------------------------%

Let us next examine the effect of the FUV radiation on limiting the minimum SFE. The red line in each panel of Fig.~\ref{fig:SFE} represents the cases with FUV feedback (i.e., Criterion 2, $\varepsilon_{\rm min,2}$). Comparing the red line to the gray line, we can evaluate the effect of the FUV feedback on top of the EUV feedback. 
The minimum SFE is defined as $\varepsilon_{\rm min} = \min(\varepsilon_{\rm min,1}, \varepsilon_{\rm min,2})$.

Fig.~\ref{fig:SFE} (a) shows that introduction of the FUV feedback does not change the SFEs in the cases with cloud mss $M_{\rm cl} = 10^4$~M$_{\odot}$; $\varepsilon_{\rm min} = \varepsilon_{\rm min,1}$. For more massive clouds with $M_{\rm cl} = 10^6$~M$_{\odot}$ (panel c), by contrast, the FUV feedback is quite important; $\varepsilon_{\rm min} = \varepsilon_{\rm min,2}$. 
For a given $\Sigma_{\rm cl}$, the minimum SFE is reduced by the inclusion of the FUV feedback by one order of magnitude, at maximum. In particular, the difference is larger at smaller surface density, $\Sigma_{\rm cl}$. In the case with intermediate mass of $M_{\rm cl} = 10^5$~M$_{\odot}$ (panel b), the resulting $\varepsilon_{\rm min}$ is only slightly (by about 10~\%) reduced by the FUV feedback effect at the lower and higher ends of $\Sigma_{\rm cl}$, i.e., $\Sigma_{\rm cl} \lesssim 100$~M$_{\odot}$pc$^{-2}$ and $\Sigma_{\rm cl} \gtrsim 400$~M$_{\odot}$pc$^{-2}$.

We also study the parameter dependencies of SFEs in the case with cloud mass $M_{\rm cl} = 10^6$~M$_{\odot}$, where the effect of the FUV feedback is the most remarkable. 
We consider the different threshold temperatures between 50~K and 300~K, and lower abundances of carbon and oxygen \citep[e.g.,][]{Cardelli96,Cartledge04}.
We find that the variations of SFEs are the most visible when the surface density is low, and the differences amount to a factor of ten at most. However, the overall trend remains the same irrespective of parameter values.

To summarize, the FUV feedback is sufficiently effective in massive and low surface density clouds. We further analyze our calculations to interpret the results in next Section~\ref{sssec:why}.

%-------------------------------------------------------------------------------------------%

\subsubsection{Interpreting Results}
\label{sssec:why}

\begin{figure*}
\begin{center}
	\begin{tabular}{c}
	\begin{minipage}{0.45\hsize}
	\begin{center}
		\includegraphics[width = \columnwidth]{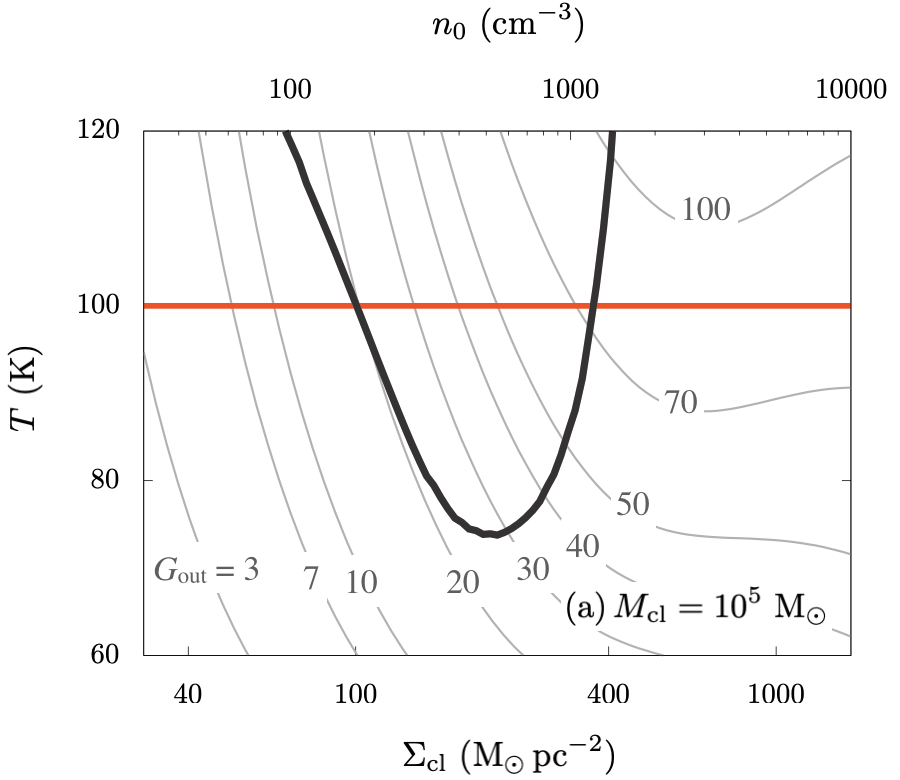}
	\end{center}
	\end{minipage}
	
	\begin{minipage}{0.45\hsize}
	\begin{center}
		\includegraphics[width = \columnwidth]{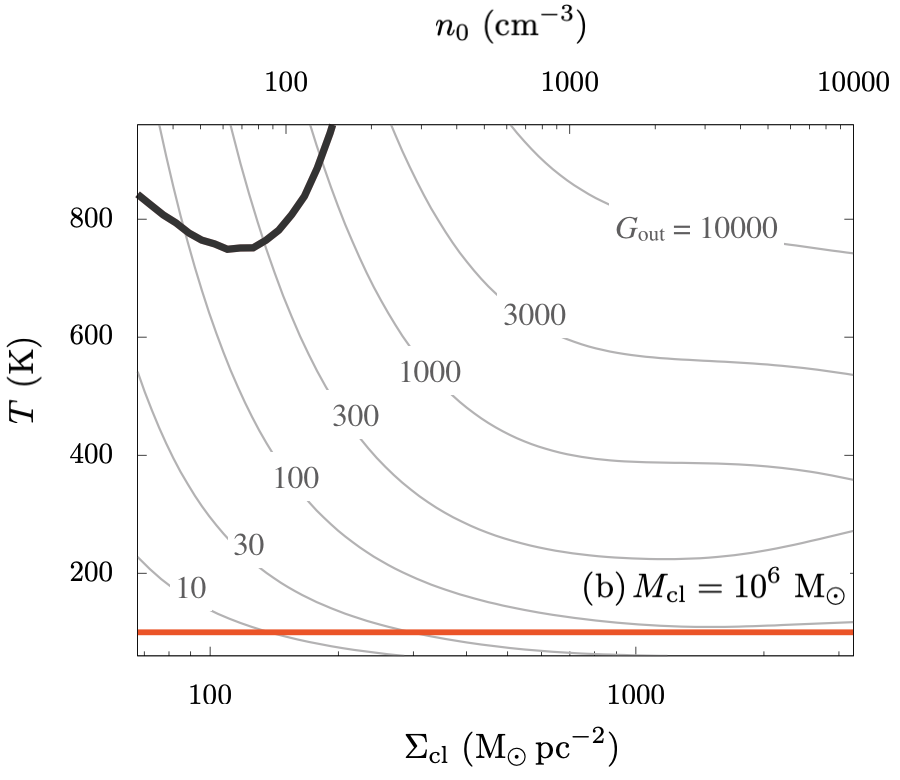}
	\end{center}
	\end{minipage}
	\end{tabular}

	\caption{
	Effects of the FUV heating in limiting the minimum SFE with $M_{\rm cl} = 10^5$~M$_{\odot}$ (panel a) and $M_{\rm cl} = 10^6$~M$_{\odot}$ (b). In each panel, the thick black line represents the gas temperature at the cloud outer edge when the minimum SFE is determined only by the EUV feedback (Criterion 1, the gray lines in Fig. \ref{fig:SFE}). In such a case, an expanding H~II region and surrounding shell just fill the whole cloud, and the "cloud edge" corresponds to the un-shocked gas just outside the shell. The red line represents the critical temperature 100~K, above which the star formation is assumed to be suppressed by the FUV feedback (Criterion 2). At $\Sigma_{\rm cl}$ for which the red curve exceeds the red line, the gas is heated up above 100~K before the shell reaches the cloud edge, meaning that the SFE should be primarily limited by the FUV feedback if included. The gray contours denote the equilibrium temperature for different values of FUV flux $G_{\rm out}$ as functions of the density. We note that, in panel (b), the vertical axis covers the much larger range of the temperature than in panel (a). 
	}
	\label{fig:n-T}
\end{center}
\end{figure*}

As shown in Section~\ref{sssec:SFE}, the impacts of the FUV feedback on limiting the minimum SFE depends on the cloud parameters such as the cloud mass $M_{\rm cl}$ and surface density $\Sigma_{\rm cl}$. Here we further look into our results to consider what causes such variations. 

%-----------------------------------------------------------------------------------%

First we investigate the case of clouds with $M_{\rm cl} = 10^5$~M$_{\odot}$. 
Since the heating in PDRs is assumed to limit the SFEs, we consider the temperature just outside of the shell, $T_{\rm out}$. The thick black line in Fig. \ref{fig:n-T}(a) shows $T_{\rm out}$ as a function of $\Sigma_{\rm cl}$ at the cloud edge $r = R_{\rm cl}$ at $t = t_{R_{\rm cl}}$, i.e., when the SFE is determined by the EUV feedback only (Criterion 1). We see that $T_{\rm out}$ has the local minimum at $\Sigma_{\rm cl} \simeq 200$~M$_{\odot}$pc$^{-2}$. 
%%%and it increases toward both the lower and higher sides. 
Since the PDR is primarily heated up via the photoelectric emission from grains, the local FUV flux $G_{\rm out}$ is a key quantity to determine $T_{\rm out}$. According to equations (\ref{eq:rcl}) and (\ref{eq:Gi}), $G_{\rm out}$ is proportional to $S_{\rm FUV} \Sigma_{\rm cl} / M_{\rm cl} \propto \varepsilon_{\rm min} \Sigma_{\rm cl}$ (neglecting dust attenuation).
It follows that $G_{\rm out}$ monotonically increases with increasing $\Sigma_{\rm cl}$, because the minimum SFE or $S_{\rm FUV}$ increases with $\Sigma_{\rm cl}$ (Fig.~\ref{fig:SFE}a). With the above facts, one may ask why $T_{\rm out}$ decreases with $\Sigma_{\rm cl}$ for $\Sigma_{\rm cl} \lesssim 200$~M$_{\odot}$pc$^{-2}$, where $G_{\rm out}$ {\it increases} with $\sgcl$.
This is explained by the nature of the [C II] line cooling, which dominates over other processes. The [C II] cooling rapidly becomes efficient with the increasing density $n$ (or $\sgcl$) for $n \ll n_{\rm cr} \simeq$ 2000~cm$^{-3}$. Such a trend is illustrated as gray lines in Fig. \ref{fig:n-T}, which show the equilibrium gas temperature as a function of density at for different values of $G_{\rm out}$ clearly show such a trend. Since the slope of the contour lines are so steep that $T_{\rm out}$ drops while $G_{\rm out}$ increases with $\sgcl$. 

%----------------------------------------------------------------------------%

Let us compare $T_{\rm out}$ with the threshold temperature for the FUV feedback, 100~K. We see that $T_{\rm out}$ exceeds 100~K in both the lower and higher sides of $\sgcl$. It suggests that the dertruction by the FUV feedback is more effective than dynamical disruption.
Since the temperature gets lower with the lower $G_{\rm out}$ at a given $\sgcl$, only the smaller $S_{\rm FUV}$ (or smaller $\varepsilon$) is enough to realize $T_{\rm out} = 100$~K. The above explains why $\varepsilon$ is reduced by the FUV feedback in the higher and lower sides of $\sgcl$ in Fig.~\ref{fig:SFE}.

%----------------------------------------------------------------------------------------%

Fig. \ref{fig:n-T}(b) shows the same plots as Fig. \ref{fig:n-T}(a) but for the cases with more massive clouds with $M_{\rm cl} = 10^6$~M$_{\odot}$, where the FUV feedback effects are more remarkable than other cases. In this case, $T_{\rm out}$ is much higher than the threshold temperature 100~K for any range of $\sgcl$. This is due to the dependence of $G_{\rm out} \propto S_{\rm FUV} \sgcl / M_{\rm cl}$ again. 
With a fixed value of $\sgcl$, $G_{\rm out}$ is larger with the higher $M_{\rm cl}$ because $G_{\rm out} \propto \varepsilon S_{\rm FUV}/M_\ast = \varepsilon \Xi_{\rm FUV}$ and $\varepsilon$ is enhanced following equation (\ref{eq:SFE_kim}).
The SFEs required to disrupt the natal cloud is much smaller than the case only with the EUV feedback.
Fig. \ref{fig:n-T}(b) also suggests that even with somewhat large threshold temperature $\lesssim 700$~K the FUV feedback should still reduce the minimum SFE $\varepsilon_{\rm min}$.

%%%%%%%%%%%%%%%%%%%%%%%%%%%%%%%%%%%%%%%%%%%%%%%%%%%%
%%%%%%%%%%%%%% Chemical Compositions %%%%%%%%%%%%%%%

\subsection{Chemical compositions of molecular cloud remnants}
\label{ssec:rem} 

\begin{figure}
\begin{center}
	\includegraphics[width = \columnwidth]{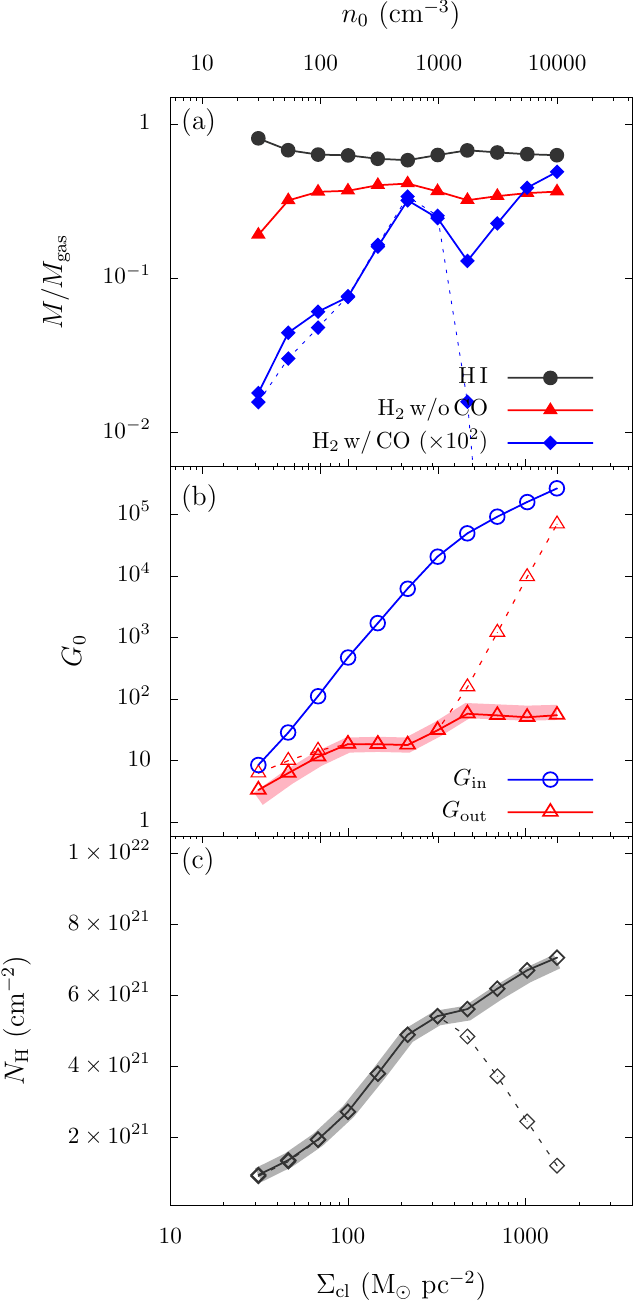}
	\caption{Chemical compositions of the gas that has not been converted into stars (molecular cloud ``remnants''', panel a) and relevant quantities (panel b and c). The same cloud mass of $M_{\rm cl} = 10^5~\msun$ is assumed for the different cloud surface densities $\Sigma_{\rm cl}$ as in Fig.~\ref{fig:SFE}. {\it Panel (a): } the mass fractions relative to the total remnant mass $M_{\rm gas} = M_{\rm cl} (1 - \varepsilon) - M_{\rm H\,II}$ for the different chemical properties: H~I (black filled circles), H$_2$ without CO (red filled triangles), and H$_2$ with CO (blue filled squares).
	{\it Panel (b): } FUV fluxes throughout the shell. The blue open circles represent the incident FUV flux at the ionization front,  and the red open triangles represent that at the preceding shock front. {\it Panel (c):} The hydrogen column density of the shell. 
In each panel, the symbols connected by the solid lines represent the cases where the minimum SFEs are limited by the EUV and FUV feedback. We also show the cases only with the EUV feedback with the thin symbols connected by the dashed lines. 
In panels (b) and (c), the thick solid lines represent the analytic evaluations of $G_{\rm out}$ and $N_{\rm H}^{\rm shell}$ by equations (\ref{eq:gout}) and (\ref{eq:nhshell}).
	}
	\label{fig:rmgas_M5}
\end{center}
\end{figure}

\begin{figure}
\begin{center}
	\includegraphics[width = \columnwidth]{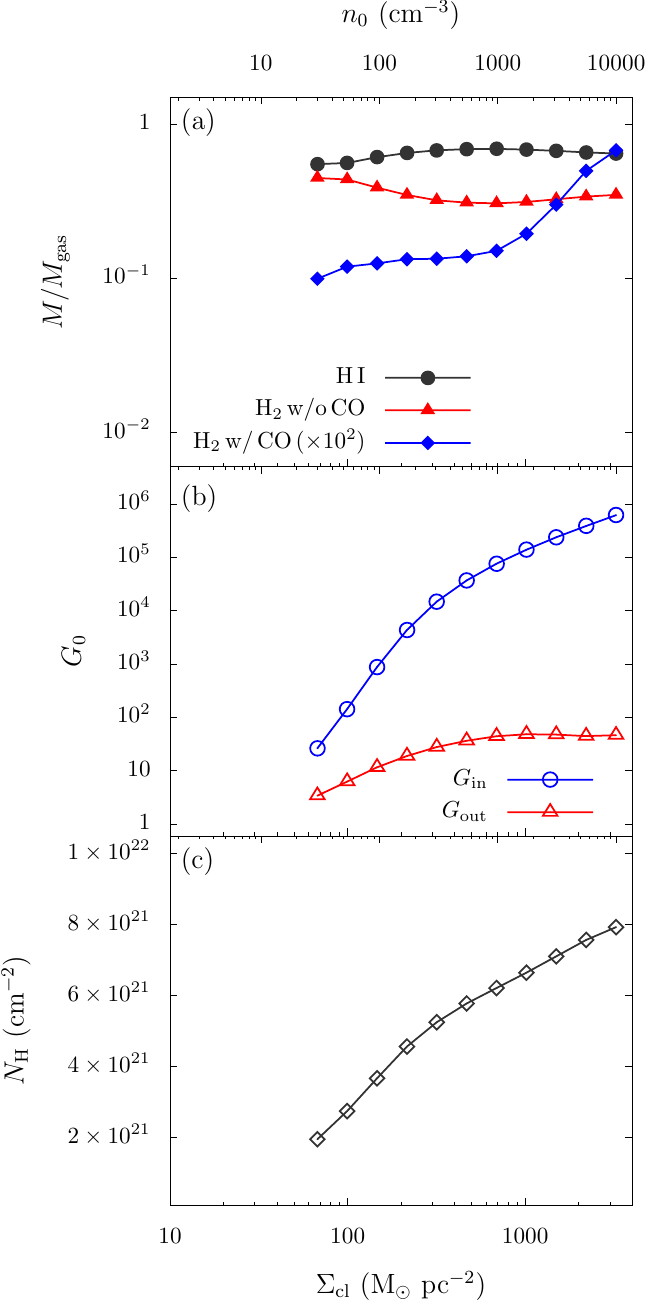}
	\caption{Same as Fig.~\ref{fig:rmgas_M5} but for the higher cloud mass of $M_{\rm cl} = 10^6$~M$_{\odot}$. In the top panel, the H$_2$-with-CO fraction for the cases only with the EUV feedback is not presented because it is far below $10^{-5}$.}
	\label{fig:rmgas_M6}
\end{center}
\end{figure}

\begin{figure}
\begin{center}
	\includegraphics[width = \columnwidth]{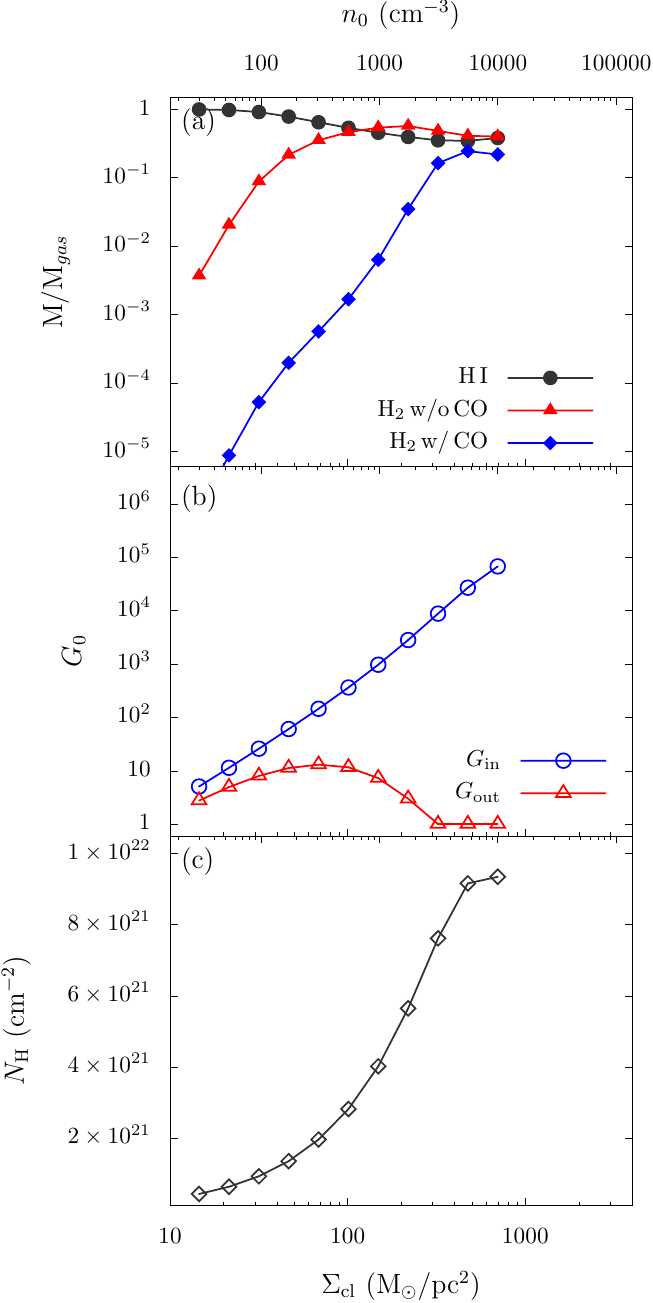}
	\caption{Same as Fig. \ref{fig:rmgas_M5} but for the lower cloud mass of $M_{\rm cl} = 10^4$~M$_{\odot}$. We here do not present the cases only with the EUV feedback unlike Figs.~\ref{fig:rmgas_M5} and \ref{fig:rmgas_M6}, because the resulting minimum SFE is exactly the same (see Section~\ref{sssec:SFE}).}
\label{fig:rmgas_M4}
\end{center}
\end{figure}

Our calculations suggest that the EUV and FUV radiative feedback from forming clusters jointly contribute to reduce the SFE of molecular clouds. In this section, we cast light on the gas that has not been used for the star formation, i.e.,  the ``remnants'' of the clouds. The cloud remnants still retain a large part of the cloud materials because the obtained SFEs are much smaller than the unity for many cases.  We here focus on the chemical compositions of the cloud remnants, which are also followed in our calculations. 

%%% 
We calculate the masses of H~I, CO-dark and CO-bright H$_2$ gases as follows:
\begin{align}
	& M_{\rm H~I} = \sum_{k = 0}^N 4 \pi r_k^2 \mu_{\rm H} \Delta N_{\rm H} x_{\rm H^0}, \\
	& M_{\rm H_2\,w/o\,CO} = \sum_{k = 0}^N 4 \pi r_k^2 \mu_{\rm H} \Delta N_{\rm H} x_{\rm H_2} x_{\rm C^+}/x_{\rm C}, \\
	& M_{\rm H_2\,w/\,CO} = \sum_{k = 0}^N 4 \pi r_k^2 \mu_{\rm H} \Delta N_{\rm H} x_{\rm H_2} x_{\rm CO}/x_{\rm C}.
\end{align}

%-------------------------------------------------------------------------------------%

First we consider the cases with the fixed cloud mass $M_{\rm cl} = 10^5$~M$_{\odot}$. 
Fig. \ref{fig:rmgas_M5} (a) presents the mass fraction of the gas with the different chemical properties as functions of $\Sigma_{\rm cl}$. The neutral and molecular hydrogens are the dominant components of the cloud remnants, and they occupy 70~\% and 30~\% of the total mass respectively. In particular, we distinguish H$_2$ molecules associated with CO molecules and those without CO. The H$_2$ gas without CO molecules is the so-called ``CO-dark'' molecular gas. 
Let us see the cases where the minimum SFE is limited by the EUV and FUV feedback (Criterion 2, solid lines). Fig. \ref{fig:rmgas_M5} (a) shows that most of the H$_2$ molecules contained in the remnants are actually CO-dark. Such a trend only has a weak dependence on $\sgcl$; the mass of the CO-dark H$_2$ gas is generally much less than 10~\% of that of the H$_2$ gas associated with CO molecules.  
This is caused by the different shielding processes of H$_2$ and CO molecules. As shown in Fig. \ref{fig:rmgas_M5}(c), the column density of the shell is roughly $N_{\rm H}^{\rm shell} \simeq 2-7 \times 10^{21}~{\rm cm}^{-2}$, corresponding to A$_{\rm V} \simeq 1-3.5$. The dust attenuation of the FUV radiation is not very efficient for such cases. In fact, Fig. \ref{fig:rmgas_M5}(b) shows that the FUV flux at the shock front $G_{\rm out}$ is several to several tens, which is high enough to photodissociate CO molecules. On the other hand, H$_2$ molecules are protected against the FUV radiation by the self-shielding effect even with the small column densities. Since the self-shielding is not available for CO molecules, which only have the small abundance, CO molecules are selectively destroyed. 

%-----------------------------------------------------------------------------------------------%

We also investigate how the above properties are altered when we only consider the EUV feedback (see the thin symbols connected with dashed lines in Fig.~\ref{fig:rmgas_M5}).
For such cases, only the quantities for $\sgcl \gtrsim 300~\sguni$ are modified. Fig.~\ref{fig:rmgas_M5}(a) shows that the amount of H$_2$ with CO molecules are further reduced for such large $\sgcl$. Fig.~\ref{fig:rmgas_M5}(c) explains it is caused by the decline of the shell column density $N_{\rm H}^{\rm shell}$. We see that $N_{\rm H}^{\rm shell}$ rather turns to decrease with $\sgcl$ for $\sgcl \gtrsim 300~\sguni$. Fig.~\ref{fig:rmgas_M5}(b) shows that $G_{\rm out}$ accordingly rises with $\sgcl$, resulting in the efficient dissociation of CO molecules.  

%------------------------------------------------------------%

The above dependence on the feedback criteria is actually well understood with the following analytic arguments.
Since the ionized gas density (at $t = t_{R_{\rm cl}}$) is given by 
\begin{equation}
	n_{\rm H\,II} = \sqrt{ \frac{3 S_{\rm EUV} f_{\rm ion}}{4 \pi R_{\rm cl}^3 \alpha_{\rm B}}},
\end{equation}
the mass of the ionized gas can be written as
\begin{align}
	M_{\rm H\,II} & = \frac{4}{3} \pi R_{\rm cl}^3 \mu_{\rm H} n_{\rm H\,II}  \nonumber \\
	& = \mu_{\rm H} \left( \frac{4 f_{\rm ion} \Xi_{\rm EUV}}{3 \alpha_{\rm B} \pi^{1/2}} \right)^{1/2} \varepsilon^{1/2} M_{\rm cl}^{5/4} \Sigma_{\rm cl}^{-3/4} \\
    & = 1.2 \times 10^4~M_{\odot} \left(\frac{\varepsilon}{10^{-2}} \right)^{1/2} \left( \frac{M_{\rm cl}}{10^5~M_{\odot}} \right)^{5/4} \left(\frac{\Sigma_{\rm cl}}{10^2~M_{\odot} {\rm pc}^{-2}} \right)^{-3/4}.
\end{align}
Since the ratio $M_{\rm H~II}/M_{\rm cl}$ depends only weakly on $M_{\rm cl}$ and $\Sigma_{\rm cl}$, we take $M_{\rm H~II} \sim 0.1~M_{\rm cl}$. Then, the shell column density $N_{\rm H}^{\rm shell}$ and FUV flux at the shock front $G_{\rm out}$ obeys the following relations
\begin{align}
	M_{\rm shell} = M_{\rm cl} (1 - \varepsilon) - M_{\rm H\,II} & \approx 4 \pi R_{\rm cl}^2 \mu_{\rm H} N_{\rm H}^{\rm shell}.
\end{align}
That is, we have
\begin{align}
N_{\rm H}^{\rm shell} & = \frac{\Sigma_{\rm cl}}{4 \mu_{\rm H}} \left( 1 - \varepsilon - \frac{M_{\rm H\,II}}{M_{\rm cl}} \right) \nonumber \\
	& \sim \frac{\Sigma_{\rm cl}}{4 \mu_{\rm H}} (0.9 - \varepsilon ), \label{eq:nhshell}\\
	G_{\rm out} & = \frac{1}{F_{\rm H}} \frac{S_{\rm FUV}}{4 \pi R_{\rm cl}^2} \exp (- \sigma_{\rm d} N_{\rm H}^{\rm shell}) \nonumber \\
	& \sim \frac{\varepsilon \Xi_{\rm FUV}}{4 F_{\rm H}} \Sigma_{\rm cl} \exp \left[ -\frac{\sigma_{\rm d}}{4 \mu_{\rm H}} \Sigma_{\rm cl} ( 0.9 - \varepsilon ) \right].
\label{eq:gout}
\end{align}
The factor of $(0.9 - \varepsilon)$ in the above equations is actually important to understand the results. Fig.~\ref{fig:SFE}(a) shows that, for $\sgcl \gtrsim 300~\sguni$, $\varepsilon_{\rm min}$ only slightly changes with whether the FUV feedback is included or not. Since $\varepsilon_{\rm min}$ is close to 0.9, however, the resulting change of $(0.9 - \varepsilon)$ is large. 
Only with the EUV feedback $(0.9 - \varepsilon)$ significantly declines, meaning that there is only little amount of the remnant gas that shields the FUV radiation. If follows that the shell column density declines for $\sgcl \gtrsim 300~\sguni$ for such cases.

%---------------------------------------------------------------------%

We have performed the same analyses as above also for the cases with the different cloud masses $M_{\rm cl} = 10^6$~M$_{\odot}$ and $10^4$~M$_{\odot}$. Fig. \ref{fig:rmgas_M6} presents the former cases with the large cloud mass $10^6~\msun$. Again, most of the hydrogen molecules contained in the cloud remnants are not associated with CO molecules (Fig.~\ref{fig:rmgas_M6}a). If we only consider the EUV feedback, we can hardly find CO molecules remained. The shell column density $N_{\rm H}^{\rm shell}$ is only less than $2 \times 10^{21}~{\rm cm}^{-2}$ (panel b), and the dust attenuation hardly contributes to reduce the FUV flux throughout the remnant gas (panel c).

%---------------------------------------------------------------------%

Similarly, Fig. \ref{fig:rmgas_M4} presents the cases with the low-mass clouds with $M_{\rm cl} = 10^4$~M$_{\odot}$. Recall that the minimum SFE does not depend on whether the FUV feedback is considered or not for this case. We see the higher fractions of H$_2$ gas associated with CO molecules than the previous cases, in particular, for $\sgcl \gtrsim 300~\sguni$ (panel a). The above analytic formulae are again useful to interpret such a variation. Since $\varepsilon_{\rm min} \ll 1$ for the current cases (see equation~\ref{eq:SFE_kim}), the factor of $(0.9 - \varepsilon)$ is just regarded as a constant. The combination of equations (\ref{eq:nhshell}) and (\ref{eq:gout}) leads to $N_{\rm H,shell} \propto \sgcl$ and $G_{\rm out} \propto \epsilon \Xi_{\rm FUV} \exp(-\sgcl)$, indicating that the FUV flux rapidly drops with increasing $\sgcl$ because the shell column density increases. Indeed, the column density $N_{\rm H,shell}$ monotonically increases with increasing $\sgcl$ (panel c). The FUV flux $G_{\rm out}$ decreases in concert, as predicted by equation (\ref{eq:gout}). For $\sgcl \gtrsim 300 \sguni$, $G_{\rm out}$ is just limited by the background value $G_{\rm out} = 1$ (panel b). The above facts suggest that the FUV radiation from the cluster is substantially attenuated by the dust grains. As a result, a certain amount of CO molecules survives, being protected against the dissociating photons.

%%%%%%%%%%%%%%%%%%%%%%%%%%%%%%%%%%%%%%%%%%%%%%%%%%

%%%%%%%%%%%%%%%%%% Discussion %%%%%%%%%%%%%%%%%%%%

\section{Discussion}
\label{sec:dis}

\subsection{Validity of thermal and chemical equilibrium}
\label{ssec:timescale}

\begin{figure}
\begin{center}
	\includegraphics[width = \columnwidth]{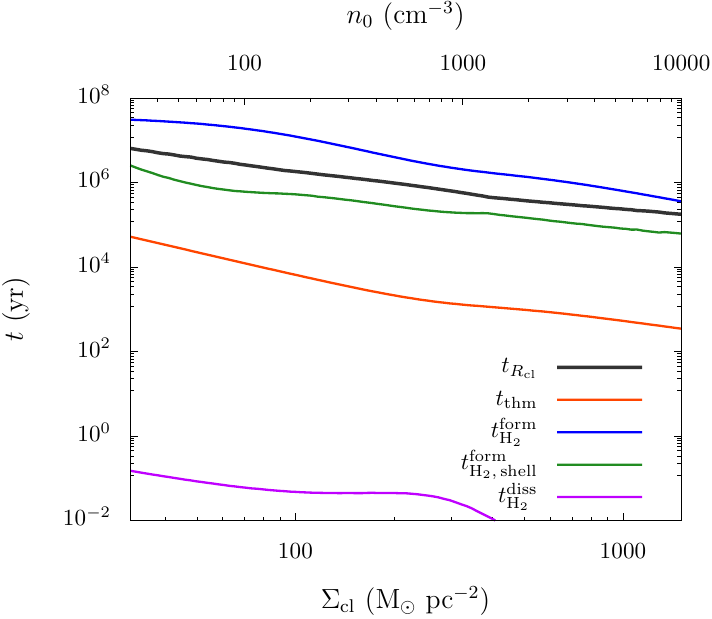}
	\caption{
Comparisons of various characteristic timescales in our calculations with $M_{\rm cl} = 10^5$~M$_{\odot}$ and different cloud surface density $\Sigma_{\rm cl}$. The snapshots when the minimum SFE is determined with our Criterion 2 are used. Presented are the shell expansion timescale $t_{R_{\rm cl}}$ (black line), cooling time at the cloud edge (at $r = R_{\rm cl}$, red line), H$_2$ formation time at the cloud edge (blue line) and on the shell (green line)
	and H$_2$ dissociation time at cloud edge (purple line).
	The average shell density is calculated by equation (\ref{eq:n_sh}.) 
	}
	\label{fig:TS_M5}
\end{center}
\end{figure}

We have assumed the thermal and chemical equilibrium in our modeling. We here examine the validity of such assumptions. In order to do that, we evaluate the timescales over which the thermal and chemical equilibrium states are achieved, $t_{\rm thm}$ and $t_{\rm chem}$.
In particular, we consider the H$_2$ equilibrium timescale $t_{\rm H_2}$ as $t_{\rm chem}$ because its formation reaction on the grain surface is slowest among the included reactions.
We calculate $t_{\rm thm}$ and $t_{\rm H_2}$ by the same method as in  \cite{KI00}.
\begin{align}
    &t_{\rm thm} = e/\Gamma,\\
    &t_{\rm H_2}^{\rm form} = x_{\rm H_2}/R_{\rm H_2}^{\rm form}, \\
    &t_{\rm H_2}^{\rm diss} = x_{\rm H_2}/R_{\rm H_2}^{\rm diss}.
\end{align}
where $R_{\rm H_2}^{\rm form}$ and $R_{\rm H_2}^{\rm diss}$   are the formation and dissociation rates of H$_2$ molecules, respectively (see equation \ref{eq:H2}).
We use the snapshots at the epochs when the expanding shell reaches the cloud edge at $r = R_{\rm cl}$, i.e., $t = t_{R_{\rm cl}}$, where the expansion timescale $t_{R_{\rm cl}}$ corresponds to the dynamical timescale.
The input parameters for calculation are $n = n_0$, 
$G_0 = G_{\rm out}$, and $N_{\rm H} = N_{\rm H}^{\rm shell}$ [Fig. \ref{fig:TS_M5} (b) and (c)].
Inside shell, in contrast,  average density is
\begin{equation}
	\bar{n} = N_{\rm H}^{\rm shell}/dR,
    \label{eq:n_sh}
\end{equation}
where $dR = r_{\rm sh} - r_{\rm IF}$ is geometrical thickness of the shell,
while $G_0$ and $N_{\rm H}$ are the same as those in the outside the shell
(this treatment is not so accurate, but is a reasonable approximation).

%---------------------------------------------------------------------------------%

In Fig. \ref{fig:TS_M5} we present the above timescales as functions of $\sgcl$ for the cases with $M_{\rm cl} = 10^5$~M$_{\odot}$. We see that all the timescales gradually decrease with increasing $\sgcl$. The dynamical timescale $t_{R_{\rm cl}} \propto R_{\rm cl}/c_{\rm s, H~II}$ decreases, because the higher $\sgcl$ is the smaller becomes the cloud size for a fixed cloud mass (equation \ref{eq:rcl}).
The chemical and thermal timescales also drop because collisions, which drive the dominant cooling and chemical processes, occur more efficiently with the higher density. The figure shows that the thermal equilibrium timescale is always much shorter than the dynamical time, thus supporting our assumption of the thermal equilibrium.

The chemical equilibrium should hold within the dense shell, which carries most of the remnant gas, since the H$_2$ formation timescale is comparable to or shorter than the dynamical time $t_{R_{\rm cl}}$ (see the green line). By contrast, the H$_2$ formation timescale is somewhat longer than $t_{R_{\rm cl}}$ at the cloud edge (see the blue line). It means that the chemical equilibrium of H$_2$ molecules may not be achieved in the un-shocked ambient medium outside of the shell by the end of the calculations. However, the clouds we consider are initially fully molecular so that the chemical equilibrium should always be a good assumption even for H$_2$, since the dissociation timescale is much shorter than the formation timescale (see the purple line).
Therefore our conclusion on the chemical composition presented in Section \ref{ssec:rem} will not change much even if we include the non-equilibrium effects.

%--------------------------------------------------------------------------------------------%

\subsection{Effects ignored}
\label{ssec:ignored}

As already mentioned in Section~\ref{sssec:one-zone}, our 1D models of the PDR use assumptions for simplicity, e.g., the optically thin fine-structure line cooling and constant dust temperature throughout a PDR. In order to examine the validity of our treatments, we have also calculated the dynamical evolution of an H~II region and surrounding PDR using a 1D radiation-hydrodynamics (RHD) code developed in \citet{HI06} for several representative cases. 
The RHD code takes the effects ignored in the semi-analytic models into account, such as the trapping effect of the line emission and variable dust temperature. We have confirmed that the simulation results show the similar overall structure of the PDR as provided by the semi-analytic models in spite of the numerous differences.  For instance, the evolution of the average density within the shell only differs by a few $\times$ 10~\% between the RHD simulations and the semi-analytic models.   

%-----------------------------------------------------------------%

Although our RHD simulations and semi-analytic models employ the same method of \cite{NL99} for the CO formation rate, there are differences in evaluating the CO photodissociation rate. 
The semi-analytic models only use the FUV intensity $G_i$, for which the dust attenuation law is given by the cross section $\sigma_{\rm d} = 10^{-21}~{\rm cm}^2 {\rm H}^{-1}$, to evaluate the CO dissociation rate.  
The RHD simulations, on the other hand, consider another FUV component only representing the CO dissociating band, for which the dust cross section is somewhat larger than the averaged value for the full FUV range $6~{\rm eV} \leq h \nu \leq 13.6~{\rm eV}$. Moreover, the RHD simulations also incorporate the effects of self- and H$_2$-shielding of CO molecules against dissociating photons \citep[e.g.,][]{vanDishoeck88}. 
The semi-analytic models thus tend to overestimate the CO photodissociation rate, ignoring these effects. In order to evaluate this effect, we have compared the simulation and model results for the case with $M_{\rm cl} = 10^5~\msun$ and surface density $\Sigma_{\rm cl} = 300$~M$_{\odot}$pc$^{-2}$ (e.g., see Figs.~\ref{fig:NH-TX_M5} for the model). As shown in Figure~\ref{fig:rmgas_M5}, the model predicts that only $\sim 0.1$~\% of the cloud remnant should be H$_2$ molecular gas associated with CO molecules. The RHD simulation run with the same setting shows that this quantity is $\sim 1$~\% at the epoch when the expanding shell reaches the cloud edge, $t \simeq 6 \times 10^5$ years since the birth of the H~II region. We interpret that such a high value in the simulation run is due to the CO dissociation rate overestimated in the model. If we ignore the effects which are not considered in the model, the simulation returns the lower value $\sim 0.03$~\%. 
We have also found that the value rapidly rises in the corresponding stage, varying by an order of magnitude in $\sim 10^5$ years. We conclude that, while there is the general trend that most of the molecular gas contained in the cloud remnants should be CO-dark, the exact amount of the CO-bright molecular gas is difficult to be accurately estimated. 
%%%
Nonetheless, it would be intriguing to investigate how the dispersing clouds are to be observed as a time sequence. For that purpose, C atoms rather than CO molecules are a more useful tracer of the CO-dark gas because of the higher abundance \citep[e.g.,][]{Li2018}. 
Coupling an extended chemistry network beyond the approximation method by \citet{NL99} with time-dependent hydrodynamics simulations should provide such predictions.

\subsection{Other stellar feedback processes}
\label{ssec:otherfb}

In order to isolate potential roles of the FUV feedback during the cloud disruption, we have employed the simple assumption on the H~II bubble expansion, i.e., that the thermal pressure excess of the photoionized gas with respect to the ambient medium drives the expansion. As briefly noted in Section~\ref{ssec:HIIdyn}, theoretical studies suggested that radiation pressure exerted on the shell affects the expansion motion \citep[e.g.,][]{KM09,Fall10,Murray10,Kim16}.
Such studies all show that the expansion is mainly driven by the radiation pressure rather than the gas pressure if $\sgcl \gtrsim 100~\sguni$, which is also confirmed by recent numerical simulations, although for turbulent clouds the transition occurs at somewhat higher $\Sigma_{\rm cl}$ \citep[e.g.,][]{Kim18}.
\citet{Kim16} have actually incorporated the effect of the radiation pressure in their model by taking $F_{\rm rad} = L/c$ as the average radiation force. We also follow the same approach as theirs to modify the temporal evolution of the shell radius given by equation (\ref{eq:rsh}). The resulting minimum SFEs for such cases are also presented by the blue dashed line in Fig.~\ref{fig:SFE}(b), for which only the EUV feedback is assumed (Criterion 1) with $M_{\rm cl} = 10^6$~M$_{\odot}$. We find that the radiation pressure effect further reduces $\varepsilon_{\rm min}$, and that its effect is more prominent for the higher $\sgcl$. 
Inversely, the FUV feedback are effective for the low surface density $\sgcl \lesssim 100~\sguni$ (Section~\ref{sssec:SFE}), for which the effect of the radiation pressure is limited.

%-----------------------------------------------------------------------------------%

Stellar winds from high-mass stars are also omitted in our models, though they have been referred to as the main driver of the bubble around a massive cluster including many O-type stars \citep[e.g.,][]{McKee84}. The dynamics of the wind-driven bubbles has been modeled assuming the spherical symmetry \citep[e.g.,][]{Weaver77}, and it is well described by an expansion law which differs from equation (\ref{eq:rsh}). Recent studies further investigate the interplay between the radiation pressure and stellar winds during the bubble expansion \citep[e.g.,][]{Rahner17,Rahner19}. Since we have focused on the FUV feedback based on the model of \citet{Kim16}, we have ignored the wind effects following their approach. 
Regarding the minimum SFEs, we have shown that the FUV feedback is effective for massive GMCs with $M_{\rm cl} \gtrsim 10^5~\msun$ (Section~\ref{ssec:SFE}). The stellar winds may affect the bubble dynamics for such cases, where the birth of massive clusters with $\gtrsim 10^3~\msun$ is supposed assuming $\varepsilon \sim 0.01$.
We have also shown that the FUV radiation produces the CO-dark gas even for the less massive clouds with $M_{\rm cl} \lesssim 10^5~\msun$ (Section~\ref{ssec:rem}). The star cluster considered is relatively small with a few O-type stars at most, for which the wind effect should be limited. 
In any rate, recent studies point out that the wind effects on the bubble expansion should be overestimated in 1D modeling. Multi-dimensional simulations show that the hot gas generated in the wind-driven bubble actually quickly leaks out through low-density channels rather than being confined \citep[e.g.,][]{Rogers13}. There are no clear observational signatures that the bubble expansion is evidently driven by the winds \citep[e.g.,][]{Lopez14}. We note that multi-dimensional effects should also affect the H~II bubble dynamics even without the wind effects, which is further discussed in Section~\ref{ssec:multiD}.

%------------------------------------------------------------------------------------%

In this paper, we have considered the stellar feedback on GMCs before the first supernova explosion occurs. As presented in Fig.~\ref{fig:TS_M5}, the dynamical timescale of an H~II bubble expansion is longer for the lower cloud surface density, $\simeq$ several $\times$ Myr for $\sgcl \lesssim 100~\sguni$. This is still shorter than the lifetime of high-mass stars that cause the supernova explosions $\sim 10$~Myr, but there may not be a long time lag.
It is interesting to speculate what happens if a supernova explosion occurs within a clouds under the stellar FUV feedback.  Since the supernova explosion add mechanical feedback on the cloud, it further contributes to reducing the SFE. Moreover, shock waves around the expanding supernova remnant sweep up the gas of the cloud being destroyed, which contains the CO-dark gas under the FUV feedback. Since the shock compression is a possible channel of the molecular cloud formation \citep[e.g.][]{II08,II09}, the CO-dark gas may be brought back into ``CO-bright'' molecular phase once the FUV radiation is somehow attenuated. 
Note that key chemical reactions producing CO molecules near the supernova remnants should differ from those in normal star-forming environments \citep[e.g.,][]{Bisbas2017}. 

%--------------------------------------------------------------%

\subsection{Inhomogeneous cloud density structure}
\label{ssec:multiD}

In our one-dimensional semi-analytic modeling, we have assumed the homogeneous density distribution within a molecular cloud. It is actually possible to relax such an assumption by improving our current model. \citet{Kim16} have also considered cases with the power-law density distributions $\rho \propto r^{-w}$ with $w < 1.5$. In general, the photoionized gas expands more rapidly with the less efficient ``trapping'' for the cloud with the steeper density gradient \citep[e.g.,][]{Franco90}. An extreme case is known as the ``champagne flow'' or ``blister-type'' H~II regions \citep[e.g.,][]{Tenorio-Tagle79}, for which the gas motion is not adequately described as the pressure-driven expanding shell, but rather as the photoevaporation where the ionized gas freely escapes from the cloud. Fully investigating the FUV feedback with such a variety of dynamical evolution is out of scope of the current work, but further studies are warranted \citep[e.g.,][]{Hosokawa07b,Geen19}. 

%----------------------------------------------------------------------------------------------------------%

In order to consider the more realistic clumpy cloud structure, one has to resort to 3D radiation-hydrodynamics numerical simulations. A number of authors in fact have conducted such simulations mostly focusing on the stellar EUV feedback
\citep[e.g.,][see also Section~\ref{sec:intro}]{Walch2012}.
%%%(Section~\ref{sec:intro}). 
Simulations by \cite{Kim18} have followed the EUV feedback againt clumpy and turbulent GMCs to drive SFEs as functions of the cloud masses and surface densities. They have confirmed the qualitative agreements with \cite{Kim16}'s model predictions, but also found that the model underestimates minimum SFEs compared to the simulation results. The simulations show that the ionized gas escapes from a cloud through low-density parts and the actual feedback is dominated by photoevaporation of surviving clumps. The FUV feedback in the clumpy medium has yet to be fully studied by similar numerical approaches \citep[e.g.,][]{Arthur11}.
Although we just have assumed that the star formation is locally quenched in a warm PDR (Section~\ref{ssec:cri}), it should be also verified with such simulations. Note that the star formation might be rather induced in a clumpy PDR because pre-existing clumps exposed to the FUV radiation would be compressed via the radiation-driven implosion \citep[e.g.,][]{Gorti02,Walch2013,Walch2015,Nakatani18}.

%%%%%%%%%%%%%%%%%%%%%%%%%%%%%%%%%%%%%%%%%%%%%%%%%%

%%%%%%%%%%%%%%%%%% Conclusion %%%%%%%%%%%%%%%%%%%%

\section{Conclusion}
\label{sec:conc}

We have developed a semi-analytic model to investigate the FUV feedback on molecular clouds, particularly effects on the thermal and chemical states of the irradiated gas.
On the basis of the previous model by \citet{Kim16}, we have solved the thermal and chemical structure of the PDR as well as the dynamical expansion of an HII region assuming spherical symmetry.
We have first evaluated the impacts of the FUV feedback on the resulting minimum SFEs supposing that the star formation is suppressed in the warm PDR where the temperature is more than a threshold value, i.e., $\sim 100$~K. We have also calculated the chemical composition of the gas that is not converted to stars, i.e., the cloud remnants, under the FUV radiation from the newborn star cluster.

%-------------------------------------------------------------------------%

Following \citet{Kim16}, we have calculated the minimum SFEs as functions of the cloud surface density $\Sigma_{\rm cl}$ for different cloud masses of $M_{\rm cl} = 10^4, 10^5, 10^6 $~M$_{\odot}$.
We argue that the FUV feedback is more effective than the pure EUV feedback caused only by the expansion of the HII regions, particularly for massive clouds with $M_{\rm cl} > 10^5$~M$_{\odot}$ and with the low surface density, $\Sigma_{\rm cl} < 100$~M$_{\odot}$ pc$^{-2}$. The minimum SFEs are reduced by the FUV feedback by no less than an order of magnitude when the star formation is assumed to be suppressed above the threshold temperature, 100~K. A key quantity to interpret such dependencies is the FUV flux at the cloud edge $r=R_{\rm cl}$ when the cloud is assumed to be disrupted by the EUV feedback, $G_{\rm out}$. If $G_{\rm out}$ is large enough, it means that the cloud is sufficiently heated up by the FUV radiation before the EUV feeedback operates, suggesting that the minimum SFE is predominantly determined by the FUV feedback. Our analyses show the scaling relation $G_{\rm out} \propto M_{\rm cl}^{1/2} \Sigma_{\rm cl}^{7/2}$, which explains why the FUV feedback is more effective with the higher $M_{\rm cl}$. 
 The same scaling suggests that $G_{\rm out}$ is rather smaller with the lower $\Sigma_{\rm cl}$ for a given cloud mass $M_{\rm cl}$, which apparently contradicts with the trend that the FUV feedback is more effective for the lower $\Sigma_{\rm cl}$. The discrepancy is explained by the fact that the [C~II] line cooling, the dominant process,  becomes inefficient sharply with decreasing $\Sigma_{\rm cl}$ (or the volume density for a fixed $M_{\rm cl}$). Owing to this, the cloud gas tends to be easily heated up even by the weak FUV radiation field. Therefore, the minimum SFE is limited primarily by the FUV feedback with the lower $\Sigma_{\rm cl}$.

%------------------------------------------------------------------%

Moreover, our analyses on the chemical compositions of the cloud remnants suggest that a large part of them are actually ``CO-dark'', except for the cases with $M_{\rm cl} = 10^4$~M$_\odot$ and $\Sigma_{\rm cl} > 300$~M$_{\odot}$ pc$^{-2}$. This is because the column densities of the cloud remnants are $2-7 \times 10^{21}~{\rm cm}^{-2}$ with the wide range of parameters $M_{\rm cl}$ and $\Sigma_{\rm cl}$. With such small column densities corresponding to $A_{\rm V} \simeq$ a few, CO molecules within the cloud remnants are not protected against the incident FUV radiation by the dust attenuation. Only hydrogen molecules survive with the efficient self-shielding effect by contrast. We have also confirmed that such a feature should be the same even for cases where the minimum SFE is primarily limited by the EUV feedback, i.e., where the stellar FUV radiation only plays a minor role in destroying the natal clouds. 
The dispersed molecular clouds are potential factories of the CO-dark gas, which returns into the cycle of the interstellar medium.

\section*{Acknowledgements}
We thank Shu-ichiro Inutsuka for fruitful discussion and comment. This work is financially supported by the Grants-in-Aid for Basic Research by the Ministry of Education, Science and Culture of Japan 
(16H05996, 19H01934: T.H.). J.-G.K. acknowledges support from the Lyman Spitzer, Jr. Postdoctoral Fellowship at Princeton University.

%%%%%%%%%%%%%%%%%%%%%%%%%%%%%%%%%%%%%%%%%%%%%%%%%%

\section*{DATA AVAILABILITY}
The data underlying this article will be shared on reasonable request to the corresponding author.



\bibliographystyle{mnras}
\bibliography{ref} % if your bibtex file is called example.bib

%\begin{thebibliography}{99}
%\end{thebibliography}

%%%%%%%%%%%%%%%%%%%%%%%%%%%%%%%%%%%%%%%%%%%%%%%%%%

%%%%%%%%%%%%%%%%% APPENDICES %%%%%%%%%%%%%%%%%%%%%

\appendix

\section{Mass-to-luminosity ratio}
\label{app:ML}

\begin{figure*}
\begin{center}
	\begin{tabular}{c}
	\begin{minipage}{0.45\hsize}
	\begin{center}
		\includegraphics[width = \columnwidth]{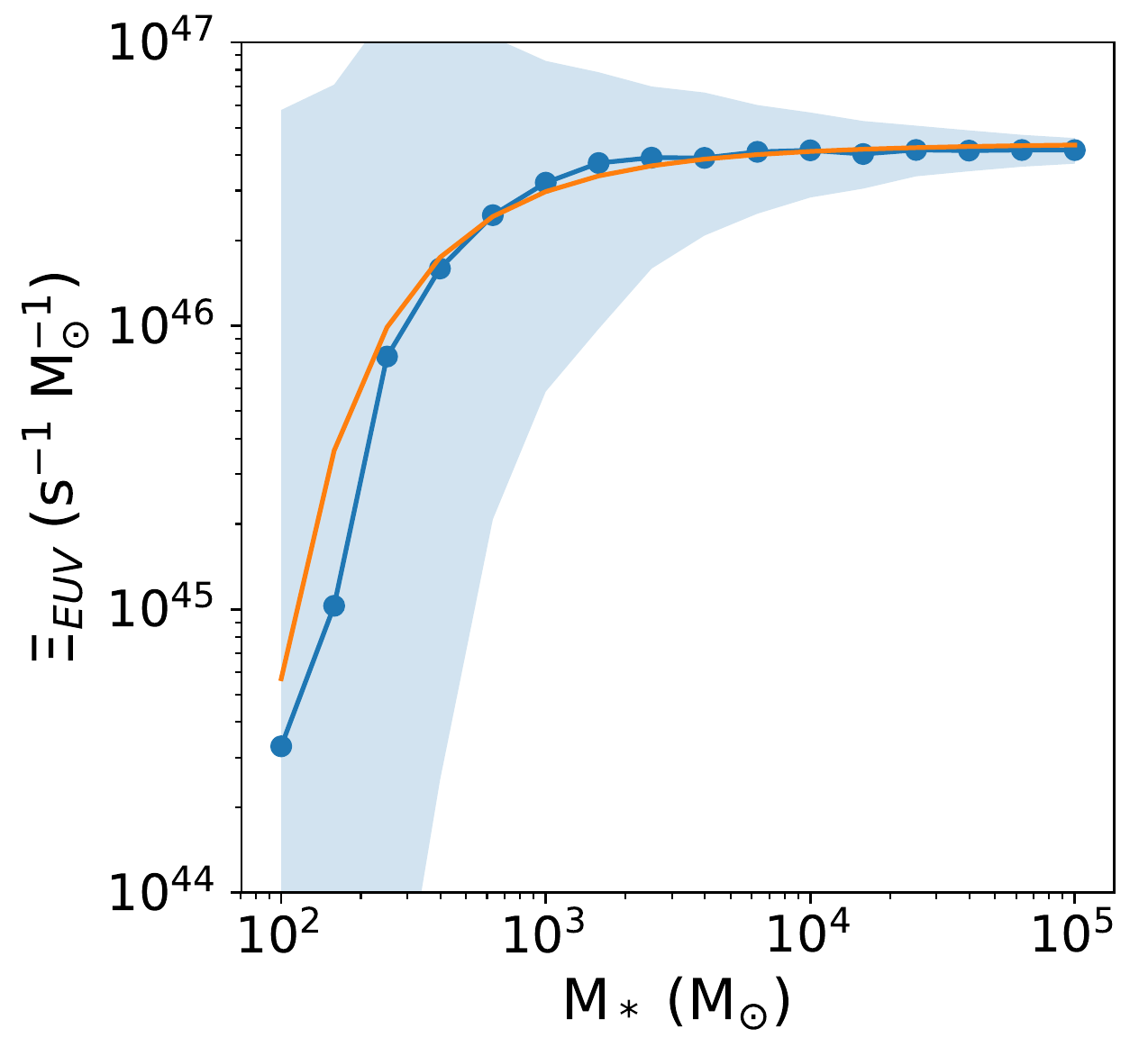}
	\end{center}
	\end{minipage}
	
	\begin{minipage}{0.45\hsize}
	\begin{center}
		\includegraphics[width = \columnwidth]{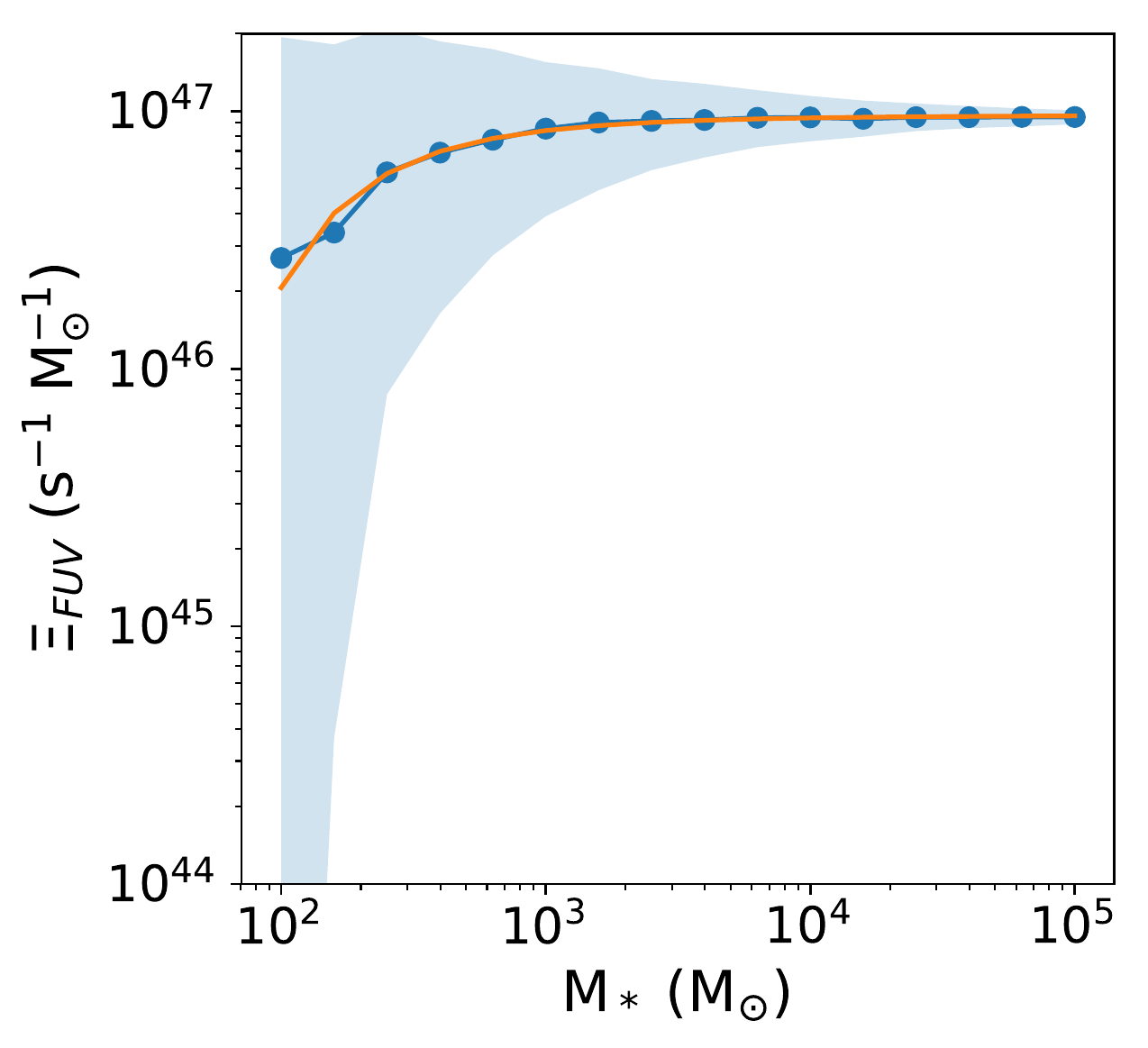}
	\end{center}
	\end{minipage}
	\end{tabular}

	\caption{
	The ratio of EUV and FUV photons emitted per unit time to stellar mass 
	$\Xi_{\rm EUV} = S_{\rm EUV}/M_{\ast}$ and $\Xi_{\rm FUV} = S_{\rm FUV}/M_{\ast}$.
	The blue line with circles represent the median value,
	while the shaded area represents the 10th to 90th percentile range from the simulation.
	Analytical fitting of the median value is showed with the orange line.
	}
	\label{fig:ML}
\end{center}
\end{figure*}

To calculate the mass-to-luminosity ratio $\Xi$ for the EUV and FUV radiation from a newborn star cluster, we use the SLUG code, a publicly available spectral population synthesis code \citep{K15}. We adopt the same settings as in \citet{Kim16}, i.e., with the IMF given by \cite{C03}, spectral synthesis model Starburst99, and stellar evolution tracks based on the Genova library. We have ran 1000 simulations for each cluster mass bin logarithmically spaced by 0.2~dex in the range of $10^2~{\rm M_{\odot}} \leq M_{\ast} \leq 10^5~{\rm M_{\odot}}$. We assume that the maximum mass of the cluster member star is 100~M$_{\odot}$. We evaluate the photon number luminosity $S_{\rm EUV}$ and $S_{\rm FUV}$ for the energy ranges of $h \nu > 13.6$~eV (EUV) and 6.0~eV $< h \nu < 13.6$~eV (FUV), respectively.

%--------------------------------------------------------------------------------%

Fig. \ref{fig:ML} presents $\Xi_{\rm EUV}$ (left panel) and $\Xi_{\rm FUV}$ (right panel) as functions of the cluster mass $M_*$. Each panel shows 10th to 90th percentile range with the blue shade and the median value with the blue circles connected by the solid line. We see that the EUV ratio $\Xi_{\rm EUV}$ rapidly decreases with decreasing the cluster mass; the values for $10^3$~M$_\odot$ are more than one order of magnitude smaller than those for $10^5$~M$_\odot$. By contrast, the FUV ratio $\Xi_{\rm FUV}$ only decreases by a factor of a few, at most, from $10^5$~M$_\odot$ to $10^3$~M$_\odot$. This is because, in comparison to the EUV cases, the less massive stars contribute more to the FUV radiation.

%----------------------------------------------------------------------------------=%

We fit the median value of $\Xi_{\rm EUV}$ and $\Xi_{\rm FUV}$ as the following analytic functions $M_{\ast}$:
\begin{align}
	& \log \left( \frac{\Xi_{\rm EUV}}{1 {\rm s}^{-1} {\rm M}_{\odot}^{-1}} \right) = \frac{46.70 \chi^6}{2.70 + \chi^6} , \\
	& \log \left( \frac{\Xi_{\rm FUV}}{1 {\rm s}^{-1} {\rm M}_{\odot}^{-1}} \right) = \frac{47.02 \chi^6}{0.92 + \chi^6} ,
\end{align}
where $\chi = \log (M_{\ast}/{\rm M_{\odot}})$. We have used these formulae in our calculations presented in the main part.

%%%%%%%%%%%%%%%%%%%%%%%%%%%%%%%%%%%%%%%%%%%%%%%%%%

% Don't change these lines
\bsp	% typesetting comment
\label{lastpage}
\end{document}